\documentclass[journal]{IEEEtranTCOM}

\usepackage{booktabs}
\usepackage{multirow}
\usepackage{fancyhdr}
\usepackage{filecontents}
\usepackage[dvipsnames]{xcolor}
\usepackage[percent]{overpic}
\usepackage{tabularx}
\usepackage{arydshln}
\usepackage{verbatim}
\usepackage{gensymb}
\usepackage{stfloats}
\usepackage{mathrsfs}

\normalsize
\usepackage{amsmath,amssymb,amsfonts}
\usepackage{algorithmic}
\usepackage{graphicx, nicefrac}
\usepackage{textcomp}
\usepackage[ruled,vlined]{algorithm2e}
\usepackage{ulem}
\usepackage{setspace} 
\usepackage{mathtools}
\usepackage{color}
\usepackage{array}
\usepackage[version=4]{mhchem}
\usepackage{float}
\usepackage{stmaryrd}
\usepackage{soul}
\usepackage{stackrel}
\usepackage{epstopdf}
\usepackage{cite}
\usepackage{subfig}
\usepackage{tikz}
\usetikzlibrary{automata,arrows,positioning,calc}
\usepackage{bigints}
\usepackage{bbm}
 \usepackage{dsfont}
\ifCLASSINFOpdf
\else
\fi

\begin{document}


\title{Achievable Information Rate Analysis in Diffusive Channels with Memory and Markov Source}

%
%
%
\author{Fardad~Vakilipoor,~\IEEEmembership{Graduate~Student Member,~IEEE},
Luca~Barletta,~\IEEEmembership{Member,~IEEE}, Stefano~Bregni,~\IEEEmembership{Member,~IEEE} and~Maurizio~Magarini,~\IEEEmembership{Member,~IEEE}
\thanks{F. Vakilipoor, L. Barletta, S. Bregni, and M. Magarini are with the Department of Electronics, Information and Bioengineering, Politecnico di Milano, I-20133, Milan, Italy e-mail: fardad.Vakilipoor@polimi.it,
luca.barletta@polimi.it,
stefano.bregni@polimi.it,
maurizio.magarini@polimi.it.}
}

\markboth{Achievable Information Rate Analysis in Diffusive Channels with Memory and Markov Source}%
{Vakilipoor \MakeLowercase{\emph{et al.}}}

\maketitle

\begin{abstract}
This paper explores the Achievable Information Rate (AIR) of a diffusive Molecular Communication (MC) channel featuring a fully absorbing receiver that counts the absorbed particles during symbol time intervals (STIs) and resets the counter at the start of each interval. The MC channel, influenced by memory effect, experiences inter-symbol interference (ISI) arising from the molecules' delayed arrival. The channel's memory is quantified as an integer multiple of the STI and a single-sample memoryless detector is employed to mitigate complexity in computing the mutual information (MI). To maximize MI, the detector threshold is optimized under Gaussian approximation of its input. The channel's MI is calculated, considering the influence of ISI, in the context of binary concentration shift keying modulation. Two distinct scenarios were considered; independent and correlated source-generated symbols, the latter modeled as a first-order Markov process. For each communication scenario, two degrees of knowledge: ISI-Aware and ISI-Unaware were considered. Remarkably, it is demonstrated that employing a correlated source enables the attainment of higher capacity. The results indicate that the capacity-achieving input distribution is not necessarily uniform. Notably, when the STI is small, corresponding to the case of strong ISI, the maximum AIR is not achieved through equiprobable symbol transmission.
\end{abstract}
\IEEEpeerreviewmaketitle
\section{Introduction}
Molecular Communication (MC) is an interdisciplinary communication paradigm that relies on particle propagation as a means of information transmission. MC has natural and artificial forms. Natural MC, which has evolved over millions of years, has great potential for investigating information exchange in biological systems. On the other hand, artificial MC is a human field that studies communication systems based on the principles of natural MC. One of the advantages of MC is its potential for use in environments where electromagnetic communication is not possible or desirable, such as in targeted drug delivery, nanomedicine, and implantable devices, for which electromagnetic radiation can be harmful or interfere~\cite{zhao2021release,rudsari2021targeted}. 

Various aspects of MC systems have been studied, including active vs. passive receivers, instantaneous vs. continuous release of molecules, and for different boundary conditions of the physical channel~\cite{jamali2019channel}. In order to better understand this novel communication paradigm, an analysis from the information theory perspective would give new insights into MC and even improve the system performance in artificial counterparts. 
\subsection{Related Literature}
\par Channel capacity serves as a fundamental metric to quantify the ability of a communication system to transmit information reliably from a sender to a receiver, as established by Shannon in his seminal work~\cite{shannon1948mathematical}. When it comes to MC, analyzing channel capacity becomes a necessary undertaking due to various factors, such as inter-symbol interference (ISI) caused by memory effects, energy constraints, slow propagation, and distinctive statistical characteristics~\cite{gohari2016information}.

One approach to investigate the capacity limits of molecular communication channels is to encode information through the timing of particle releases, as explored extensively by Rose \textit{et al.}~\cite{rose2016inscribed}. First, they illustrated that any MC channel can be contextualized either from the timing perspective or the type of particles. Then, they mainly focused on obtaining upper and lower bounds on the capacity of MC timing channels. Lastly, they applied their theory obtained from the particle counting process to DNA and protein sequences. In another study~\cite{farsad2018capacity}, an MC timing channel is introduced, where particles decay after a finite time, and upper and lower bounds on the associated capacity are derived.

Another approach in MC involves encoding information based on the number of particles released at the transmitter and decoding based on the number of received particles during the symbol time interval (STI)~\cite{farsad2020capacities}.

Early investigations into MC channel capacity in the presence of particle intensity considered receivers that do not interact with information particles (IP), which are referred to as transparent receivers~\cite{pierobon2012capacity}. However, in practical scenarios, receivers commonly engage in a reaction process, binding with the IP through natural processes.

The concentration-based channel, coupled with a ligand receptor-equipped receiver, has been the subject of investigation in previous works, such as those by Einolghozati \textit{et al.}~\cite{einolghozati2011capacity} and Tahmasbi \textit{et al.}~\cite{tahmasbi2015capacity}. These studies employ a Markov chain model to capture the reception of molecules by ligand receptors and analyze the channel capacity in terms of bits per channel use. Furthermore, the findings from these studies have been extended to multiple access channels in~\cite{aminian2015capacity}.

The binding process can be linked to the concentration of interacting particles and, equivalently, to the number of particles absorbed by the receivers. Consequently, recent efforts have been made to evaluate the capacity and establish bounds for diffusive MC channels with fully absorbing (FA) receivers. In work by Ghavami~\textit{et al.}~\cite{ghavami2015information}, the capacity of a 1D diffusive channel with an advection term was examined. Initially, they ignored the consideration of ISI. Subsequently, a memory length equivalent to two symbol intervals was introduced, thereby accommodating the effects of ISI. The authors proceeded to illustrate and examine the metrics of capacity per channel use and capacity per unit of time. In one study~\cite{ratti2020upper}, upper and lower bounds on the channel capacity were derived, assuming Poisson and Gaussian models to characterize the statistical properties of the received signal.

In another study~\cite{ratti2021bounds}, the received signal was modeled as a Poisson random variable (RV), and bounds were determined for the constrained capacity of a diffusive MC system employing concentration shift keying (CSK) as the modulation scheme. The lower bound was derived from the mutual information (MI), calculated as the difference between the marginal entropy of the output and the conditional entropy of the output given the input. On the other hand, the upper bound was derived from the dual expression for channel capacity.

In a different investigation~\cite{liu2020channel}, the channel capacity was evaluated for various reaction rates of the absorbing receiver, assuming a uniform bit probability distribution, although an optimal input distribution for bit transmission would be expected in capacity analysis. Moreover, this work considered the threshold of the memoryless detector as a predefined constant, which may not ensure optimal detection and accurate computation of the maximum MI.

\subsection{Motivation and Contribution}

In this paper, we consider an FA receiver that, under the assumption of perfect symbol synchronization, counts the number of particles absorbed along each STI and resets the counter at the beginning of the next interval~\cite{ghavami2015information}. The MC channel introduces a memory effect, due to the delayed arrival of molecules, and thus ISI. Hence we are dealing with a channel with additive memory property. The term additive indicates that the particles' delayed arrival can result in the incremental accumulation of the absorbed particles.
From the statistical perspective, the received signal in each STI follows a multinomial distribution. However, given the settings and channel characterizations, we demonstrate that it can be approximated as Gaussian.

The reset counting mechanism that we have considered at the receiver side of the channel is a concept not confined solely to this context. In the realm of electronics, analogous functionalities are realized through sample-and-hold circuits. Brain neurotransmitters perform a similar task. Through reuptake mechanisms, the transporter proteins are responsible for removing neurotransmitters from the synaptic cleft, resetting their concentration and terminating their signaling effects~\cite{mukherjee2002regulation}. Observing that in nature and electronic circuits, we felt motivated to consider such receiving mechanism and investigate the Achievable Information Rate (AIR), when the channel impulse response (CIR) varies with the transmission rate.

To mitigate the computational complexity associated with considering all possible combinations of previously transmitted symbols in the calculation of MI, we estimated the channel memory length in terms of integer multiples of STIs. For our analysis, we employed a single-symbol memoryless detector, although it is worth noting that a multi-symbol receiver may yield better performance due to the strong memory effect present in the diffusive channel. Hence, we refer to our capacity calculation as the \emph{memoryless capacity}. Previous works mentioned in this paper assumed a fixed threshold to detect the received signal. However, we optimize the threshold through a brute-force algorithm to find the memoryless capacity. Investigating the capacity regardless of a fixed predefined value of threshold allows us to better perceive the characterizations of the channels with memory with respect to the input distributions. We believe that under the optimum threshold, we can compare different scenarios in terms of bit rate per unit of time.

This paper undertakes an exploration of MI under various distinct scenarios. Initially, we contemplate a situation wherein a correlated source is aligned with a receiver possessing an awareness of preceding symbol transmissions (ISI-Aware). Subsequently, we assess the same source type but in the absence of any prior knowledge regarding past transmitted symbols (ISI-Unaware). Subsequent to this, we transition to an independent source and proceed to evaluate MI within analogous contextual configurations (\textit{i.e.,} ISI-Aware and ISI-Unaware).

The correlated source is modeled as a first-order Markov source with time-invariant transition probabilities. The capacity and associated input distributions are determined for each source type. Our findings demonstrate that, compared to independent sources, correlated sources can achieve higher capacity. This is primarily attributed to the degree of freedom that correlation offers in avoiding the consecutive transmission of identical symbols. This strategy proves particularly advantageous when the STI is short (\textit{i.e.}, high symbol transmission rate), resulting in increased ISI. Notably, we also establish that as the STI increases (\textit{i.e.}, reduced ISI), the same capacity can be achieved regardless of the source type or knowledge of previously transmitted symbols. Additionally, our research reveals a perhaps counter-intuitive observation: in scenarios involving fast symbol transmission rates, avoiding the transmission of $``1"$s (\textit{i.e.}, not releasing IPs) is not the sole optimal solution. Conversely, we demonstrate that adjusting the input distribution to allow a higher number of $``1"$s (\textit{i.e.,} releasing IPs) to be transmitted can also be a viable strategy, offering a compromise in terms of AIR. Ultimately, we believe that the superior performance of correlated sources in terms of AIRs, shown in this work, gives an insight into how to design codes for the molecular diffusive channel and, more in general, for additive Gaussian channels whose variance depends on the transmitted symbols.

We would like to point out that our approach and methodology are valid, not necessarily in MC studies. Actually, it is applicable to any channel with additive memory property under Gaussian statistics when the source is either independent or correlated.

\subsection{Outline}
\par The paper is structured as follows: Sec.~\ref{sec:system_model} introduces the system model, including the calculation of memory, as well as the CIR. Furthermore, the suitability of the Gaussian approximation for channel modeling is discussed, followed by an examination of the transition probabilities of the channel. In Sec.~\ref{sec:capacity}, we provide a detailed explanation and formulation of the memoryless channel capacity, AIR, and MI for both the independent and correlated sources. For each source type, two MIs are derived, corresponding to the ISI-Aware and ISI-Unaware scenarios. Section~\ref{sec:simulation_results} presents numerical results illustrating the capacity and AIR for the four distinct scenarios, considering various STIs and input probabilities. Finally, Sec.~\ref{sec:conclusion} concludes the paper by offering final remarks.
\subsection{Notations}
\par 
The RVs are represented by uppercase italic letters ($X$), while their realizations are denoted by lowercase italic letters ($x$). The vector $(x_{r},\ldots,x_{v})$ is expressed as $x_{r}^{v}$. Specifically, the presence of a superscript indicates that the variable is a vector, while the subscript indicates the index of the first element, and the superscript indicates the index of the last element in the vector. If there is only a subscript, it denotes a single variable with the corresponding index. Additionally, the joint probability of the vector $(x_{r},\ldots,x_{v-1})$ and $x_{v}$ can be written as $P_{X_{r}^{v-1},X_{v}}(x_{r}^{v-1},x_{v})$$\,=\,$$P_{X_{r}^{v}}(x_{r}^{v})$. The Hamming weight operator applied to a binary vector $x_{r}^{r+n}\in\{0,1\}^{n+1}$ is denoted as $w_{H}(x_{r}^{r+n})$, which counts the number of occurrences of $``1"$ in vector $x_{r}^{r+n}$. The operator $\{f\}^{+}$ is defined as the $\max \{0,f\}$. The $Q$ function and complementary error function are defined as
\begin{equation}
    Q(z) = \frac{1}{2}\mathrm{erfc}\left ( \frac{z}{\sqrt{2}} \right )= \frac{1}{\sqrt{2\pi}}\int_{z}^{\infty} e^{-\frac{y^2}{2}}dy, \quad z \in \mathbb{R}~.
\end{equation}
The binary entropy function $H_{2}:\: [0,1] \rightarrow [0,1]$ is defined as
\begin{equation}
    H_{2}(x) \triangleq -x\log_{2}(x)-(1-x)\log_{2}(1-x)~.
\end{equation}
\section{System Model and Analysis}\label{sec:system_model}
In this section, we undertake a comprehensive characterization of the transmitter, receiver, and the propagation dynamics of the IPs. Subsequently, we proceed to quantify the memory of the system, representing it as an integer multiple of the STI. Following the characterization of the CIR, we proceed to model the received signal and approximate its statistical behavior by employing a Gaussian distribution.
\subsection{Propagation Aspects}
This work considers a communication system made of a point transmitter, a diffusion-based channel, and an FA spherical receiver. At the beginning of each STI of duration~$T_{\mathrm{sym}}$ where $``1"$ is sent, the transmitter sends a pulse corresponding to the instantaneous release of $N_{\mathrm{T}}$ IPs. The receiver counts the number of particles absorbed within each STI and resets the counter at the beginning of the next interval. We believe that this mechanism is not far from reality~\cite{upadhyay2019emerging}. The IPs diffuse through the medium between transmitter and receiver with constant diffusion coefficient $D~[\mu \mathrm{m}^2/\mathrm{s}]$. In practice, the value of $D$ depends on the temperature, viscosity of the fluid, and the Stokes’ radius of the molecule~\cite{tyrrell2013diffusion}.

\par The receiver's absorption property stems from the reaction between receiver and IPs. In effect, the counting process is tantamount to measure the concentration of desired particles at the receiver, resulting from the interaction between its surface and particles. In a biological environment, enzymes can be secreted by the receiver to eliminate effects resulting from past reactions, thus enabling resetting~\cite{awan2017improving}.
\par The propagation of diffusive particles is governed by the Fick's second law~\cite{crank1975mathematics}, which relates the time derivative of the flux to the Laplacian of the concentration of molecules $c\left(d,t\right)$ at a given distance $d$ and time $t$, as
\begin{equation}
    \frac{\partial c \left( d,t \right)}{\partial t} = D \nabla^2c
    \left( d,t \right). \label{eq:2nd Fick}
\end{equation}
The initial and boundary conditions of~\eqref{eq:2nd Fick} vary depending on the MC system characteristics. Authors in~\cite{schulten2000lectures} specified the boundary and initial conditions for an impulsive release of molecules, an unbounded environment, and an FA spherical receiver. They obtained the expression for the hitting rate of molecules onto the receiver surface, as a function of the distance~$d$ between the transmitter and the center of the receiver with radius $R$ at time~$t$. Then, assuming the independent random movement of the particles and the homogeneity of the medium, they derived the expected cumulative number of absorbed particles as 
 \begin{equation}
 N(t) = \frac{N_{\mathrm{T}}R}{d} \mathrm{erfc}\left ( \frac{d-R}{2\sqrt{Dt}} \right )~.\label{eq:CNAP}
 \end{equation}
\par In our study, we consider a binary concentration shift keying (BCSK) modulation, where IPs release corresponds to $``1"$ and no release corresponds to $``0"$. At the receiver, the number of absorbed particles is counted and reset at the beginning of the next interval. At the end of each STI, the receiver returns a single sample, representing the total number of particles absorbed during that interval. Assuming that the receiver resets the counter right at the beginning of STIs (\textit{i.e.} perfect synchronization between transmission and reset intervals at the receiver), we expect that the receiver observation changes by varying the duration of the STI $T_{\mathrm{sym}}$. 

To compute the MI, we need to calculate the probability that particles hit the receiver. Since the total number of released particles is $N_{\mathrm{T}}$, if the counter has not been reset between the initial time of release until time $t$, the probability that a particle hits the receiver at time $t$ is $N(t)/N_{\mathrm{T}}$. If the counter is reset, instead, the probability that a particle released at $t$$\,=\,$$0$ hits the receiver within the $i$th STI is 
\begin{equation}
    h_{i} = \frac{N(iT_{\mathrm{sym}})-N((i-1)T_{\mathrm{sym}})}{N_{\mathrm{T}}}~, \label{eq:pi}
\end{equation}
because a particle that has been absorbed at any time $t$$\,<\,$$(i$\,$-$\,$1) T_{\mathrm{sym}}$ does not have a second chance to hit the receiver.
\subsection{Memory Duration Characterization}
\par When studying slow diffusive communication, it is important to quantify the effect of channel memory. To compute MI and transition probabilities between input and detected output, we need to account for all possible combinations of the preceding symbol sequence. If channel memory spans $M$ symbols, there are $2^{M}$ different possible sequences to consider. The memory length depends on the transmission rate of symbols. In our model, it should be as small as possible, because evaluating $2^{M}$ combinations can make computation impractical. Moreover, due to the differential nature of~\eqref{eq:pi} and asymptotical convergence of~\eqref{eq:CNAP}, the probability of a particle being absorbed a long time after release eventually tends to $0$. 

\par To obtain an estimate $M$ of the \emph{effective memory length} in terms of STIs, being not unnecessarily long or so short to miss the effect of the released particles, we define
\begin{equation}
    M = \left\lceil \frac{T_{\alpha}}{T_{\mathrm{sym}}} \right\rceil~,
\end{equation}
where $T_{\alpha}$ is the time required to reach some negligible hitting probability~$\alpha$, as given by
\begin{equation}
    \frac{R}{d}\left ( \mathrm{erfc}\left ( \frac{d-R}{2\sqrt{D(T_{\alpha}+T_{\mathrm{sym}}))}} \right ) -\mathrm{erfc}\left ( \frac{d-R}{2\sqrt{DT_{\alpha}}} \right ) \right )= \alpha~.\label{eq:alpha_criterion}
\end{equation}
Note that~\eqref{eq:alpha_criterion} is a transcendental equation with unknown~$T_{\alpha}$. We do not know an explicit solution for such equation. Hence, we solve it numerically by the \textit{regula-falsi} method~\cite{conte2017elementary}.

\par
\begin{figure}
    \centering
    \includegraphics[width=1\columnwidth]{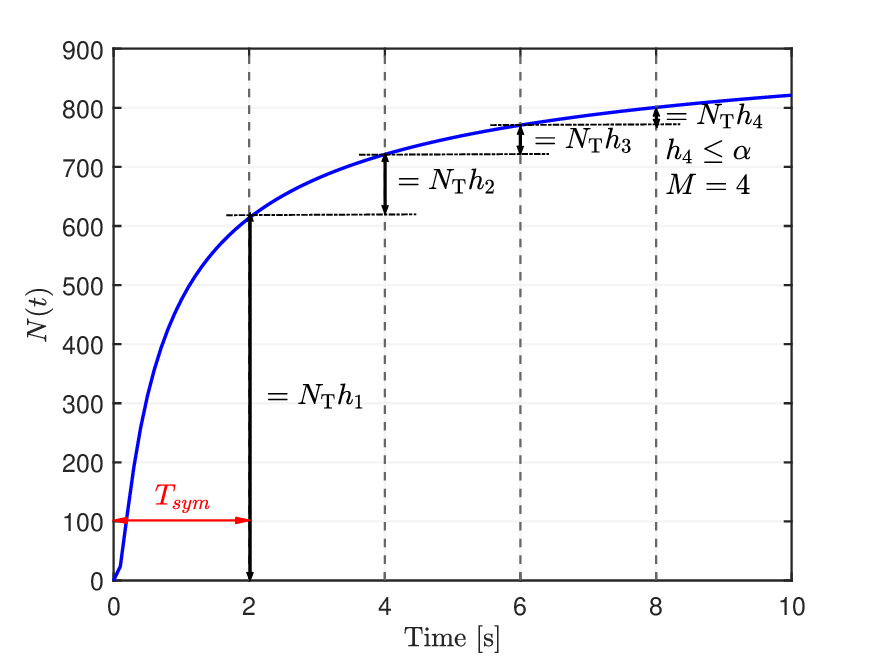}
    \caption{Expected cumulative number of absorbed particles over time without resetting (blue curve) (system parameters as in Tab.~\ref{tab:param}, $T_{\mathrm{sym}}$$\,=\,$$2$~s, $M$$\,=\,$$4$). The expected number of absorbed particles within each interval, when the counter is reset, is highlighted by vertical double arrows.}  
    \label{fig:CNAP}
\end{figure}

\begin{figure*}[htp]
    \centering
    \includegraphics[width=.325\textwidth]{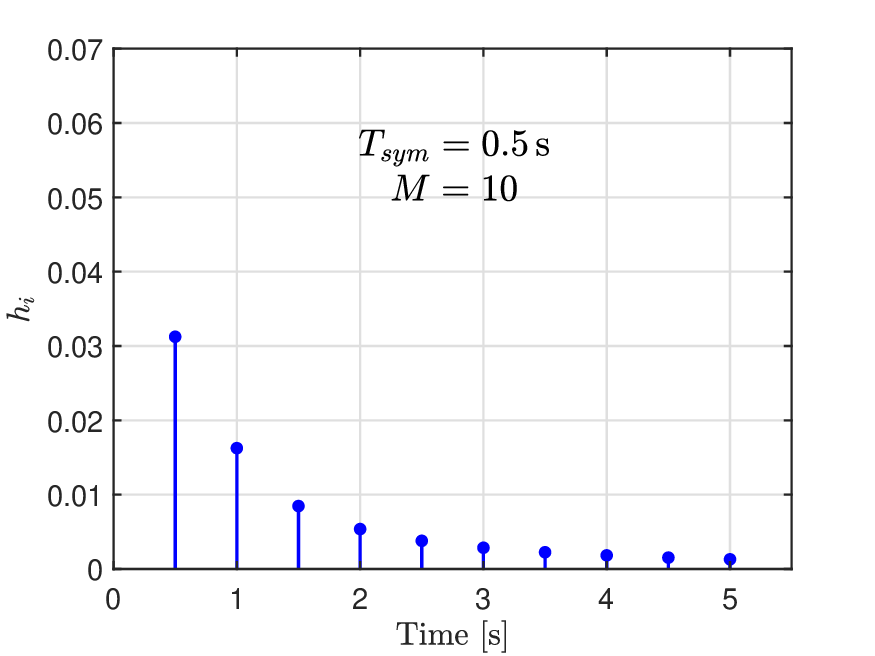}
    \includegraphics[width=.325\textwidth]{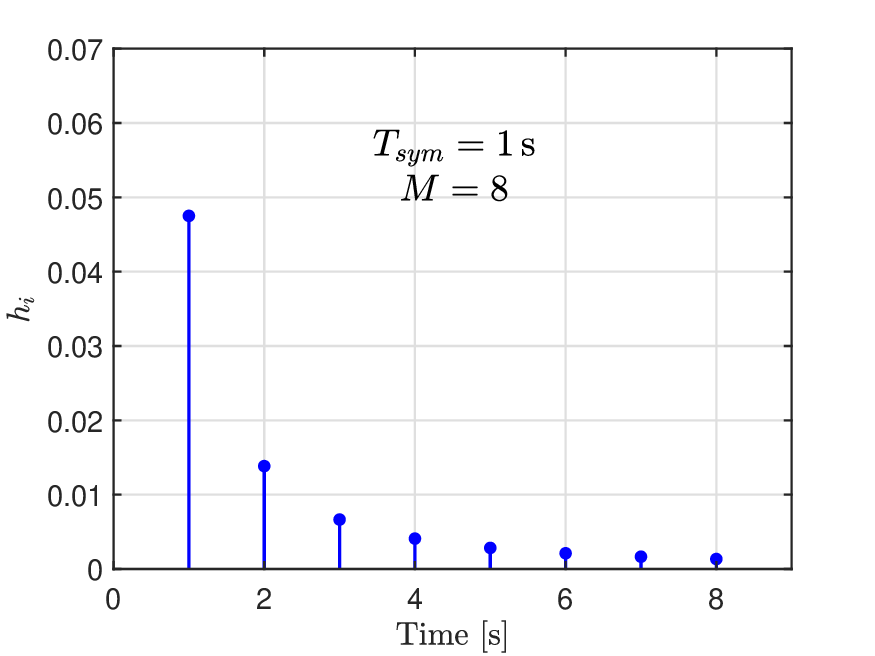}
    \includegraphics[width=.325\textwidth]{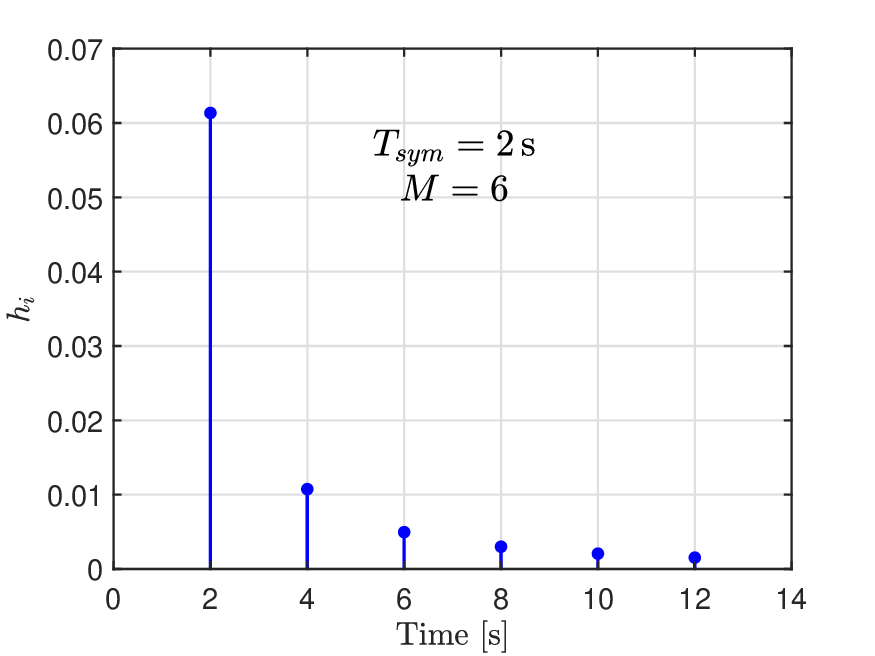}
    \caption{Distribution~\eqref{eq:pi} of the probability that particles are absorbed by a resetting receiver within the $i$th interval for $T_{\mathrm{sym}}$$\,=\,$$0.5, 1, 2$~s. The channel memory length, resulting from \eqref{eq:alpha_criterion} for $\alpha$$\,=\,$$0.001$, increases with $T_{\mathrm{sym}}$ when measured in time units ($T_{\alpha}$), but decreases in terms of STIs ($M$).}
    \label{fig:alpha_memory}
\end{figure*}

 For example, Fig.~\ref{fig:CNAP} plots the expected cumulative number of absorbed particles over time without resetting the counter (blue curve), computed by~\eqref{eq:CNAP} for a diffusive MC system modeled as above, with parameters set as in Tab.~\ref{tab:param}, and with STI $T_{\mathrm{sym}}$$\,=\,$$2\,$s. Here, the memory length results in $M$$\,=\,$$4$. The expected number of absorbed particles within each STI, resetting the counter at its beginning, is also highlighted as the difference of values at interval boundaries.
On the other hand, Fig.~\ref{fig:alpha_memory} plots the distribution~\eqref{eq:pi} of probability $h_{i}$ that particles are absorbed by a resetting receiver within the $i$th interval for different values of $T_{\mathrm{sym}}$. We observe that the memory length resulting from~\eqref{eq:alpha_criterion} increases with $T_{\mathrm{sym}}$ when measured in time units ($T_{\alpha}$), but decreases in terms of STIs ($M$). The vector $h_{1}^{M}$$\,=\,$$(h_{1},\ldots,h_{M})$ represents the CIR of the system.
\subsection{Gaussian Approximation}
\par The received signal at the $i$th STI $\mathtt{R}_{i}$ consists of the number of particles released for the $i$th transmitted symbol $\mathtt{C}_{i}$ as well as of those released for previous symbols and absorbed within the current interval $\mathtt{P}_{i}$. We consider also an environment external noise $\mathtt{E}$, due to random factors that increase or reduce the number of particles that the receiver counts in any interval. For example, a negative value of $\mathtt{E}$ expresses the effect of extraneous molecules that unbind absorbed IPs. As obvious, $\mathtt{E}$ is independent of $\mathtt{C}_{i}$ and $\mathtt{P}_{i}$. In conclusion, the observation at the $i$th interval is the superposition of the current signal, of previously transmitted symbols and of external noise, that is,
\begin{equation}
    \mathtt{R}_{i} = \mathtt{C}_{i} + \mathtt{P}_{i} + \mathtt{E}~.
    \label{eq:N_r}
\end{equation}
Next, we want to show that we can use a Gaussian model to describe the randomness in the number of absorbed particles in STIs. Due to the nature of the absorption phenomenon and the temporal correlation of the absorption in different STIs, the number of particles absorbed in each STI follows the multinomial distribution. Consider each STI as a bin. Hence, we have $M$ bins that correspond to the STIs in which a particle can reach the receiver. There is also an extra bin that represents the scenario of a particle that has not been absorbed. So in the end, for our statistical model, there are $M+1$ bins. We can write the probability of a particle falling into the $i$th of the first $M$ bins as $h_{i}$ and the probability corresponding to the last extra bin as $h_{M+1}$$\,=\,$$1-\sum_{i=1}^{M} h_{i}$. Let $N_{1}^{M+1}$$\,=\,$$(N_1, \ldots, N_{M+1})$ be the number of particles that fell into each bin. Then $N_{1}^{M+1}$ is multinomial-distributed over $N_{\mathrm{T}}$ trials and bin probabilities $h_{1}^{M+1}$. We can compute the entries of the  covariance matrix of $N_{1}^{M+1}$ as follows 
\begin{equation}
    \mathrm{Var}(N_i) = N_{\mathrm{T}}h_{i}(1-h_{i})~,\qquad i\in\{1,\ldots,M+1\}~,
\end{equation}
\begin{equation}
    \mathrm{Cov}(N_i,N_j) = -N_{\mathrm{T}}h_{i}h_{j}~,\qquad (i\neq j)~.
\end{equation}
If $h_{i}$ and $h_{j}$ are much smaller than $1$, then we have
\begin{equation}
    \frac{\mathrm{Var}(N_i)}{|\mathrm{Cov}(N_i,N_j)|}=\frac{1-h_{i}}{h_{j}}\gg 1~.\label{eq:dominantDiag}
\end{equation}
By Central Limit Theorem (CLT), for $N_{\mathrm{T}}$ sufficiently large, the random vector $N_{1}^{M+1}$ is approximately Gaussian distributed. However, our focus is on characterizing the joint distribution of the first $M$ bins: so the vector $N_{1}^{M}$ is approximately Gaussian distributed with mean vector $N_{\mathrm{T}}h_{1}^{M}$ and covariance matrix that is approximately diagonal thanks to~\eqref{eq:dominantDiag}. This is equivalent to having virtually independent and Gaussian-distributed marginals as follows
\begin{equation}
N_i\sim\mathcal{N}\Big(N_{\mathrm{T}}h_{i},N_{\mathrm{T}}h_{i}(1-h_{i})\Big),\qquad i\in\{1,\ldots , M\}.
\end{equation}
In practice, to be this approximation valid, the probability that the Gaussian distribution generates negative values should be negligible. That is, the model parameters should be chosen to have the mean $\mu$ and standard deviation $\sigma$ of the Gaussian distribution satisfying \textit{e.g.}
$\mu>3\sigma$, which implies
\begin{equation}
    \frac{N_{\mathrm{T}}h_{i}}{1-h_{i}}>9~.\label{eq:valid_apx}
\end{equation}
\par One of the necessary parts of a communication system is the detector that attempts to recover the transmitted symbol from the received signal. In this paper, we consider a memoryless binary detector with the rule
\begin{equation}\label{eq:detect}
 \hat{S}_{i}=\begin{cases}
    1 & \text{if $\mathtt{R}_{i}\geq\tau$},\\
    0 & \text{otherwise}.
  \end{cases}
\end{equation}
Note that, for each realization of the channel, such as input symbol distribution, and STIs, we always look for the threshold $\tau$ that maximizes the MI.
\par Let $s_{i}$$\,\in\,$$\{0,1\}$ denote the transmitted symbol associated with the $i$th STI and $g(\omega;\mu,\sigma^2)$ be a Gaussian probability density function (pdf) with mean $\mu$ and variance $\sigma^2$, where $g(\omega;0,0)$$\,=\,$$\delta(\omega)$ is the Dirac delta function.
Then, the pdf of the current signal conditioned on a specific realization of the current transmitted symbol ($S_{i}$$\,=\,$$s_{i}$) can be written as
\begin{equation}
    f_{ \mathtt{C}_{i}|S_{i}=s_{i}}(\omega) = g\Big(\omega;s_{i}N_{\mathrm{T}}h_{1},s_{i}N_{\mathrm{T}}h_{1}(1-h_{1})\Big)~.
\end{equation}
Let the vector $s_{i-M+1}^{i-1}$$\,\in\,$$\{0,1\}^{M-1}$ be a realization of $(S_{i-M+1},\ldots,S_{i-1})$, that is the $M$$\,-\,$$1$ symbols preceding the $i$th interval. The conditional pdf of particles released in the past $M$$\,-\,$$1$ intervals and absorbed within the $i$th  interval, given the sequence of preceding symbols, is
\begin{align}
    \begin{split}
    &f_{ \mathtt{P}_{i}|S_{i-M+1}^{i-1}=s_{i-M+1}^{i-1}}(\omega) =\\& \hspace{0.3cm} g \bigg(\omega;N_{\mathrm{T}}\sum^{M}_{j=2}s_{i-j+1}h_{j}, N_{\mathrm{T}}\sum^{M}_{j=2}s_{i-j+1}h_{j}(1-h_{j})\bigg)~.
    \end{split}
\end{align}
The external noise is assumed to follow a time-independent Gaussian distribution with pdf
\begin{equation}
    f_{ \mathtt{E}}(\omega) = g\Big(\omega;\mu_{ \mathtt{E}},\sigma_{ \mathtt{E}}^{2}\Big).
\end{equation}
To obtain the conditional pdf of the received signals, two pdfs need to be considered first. The first is the pdf of the number of particles received in the $i$th interval but released in previous intervals given the previously transmitted symbols, including the environment noise, \textit{i.e.} $ (\mathtt{P}_{i}+ \mathtt{E})|s_{i-M+1}^{i-1}$. The second is the pdf of particles received in the $i$th interval, released in the current and previous intervals given the previously transmitted symbols and the current symbol, including the external noise, that is $ (\mathtt{C}_{i}+ \mathtt{P}_{i}+ \mathtt{E})|s_{i-M+1}^{i}$. As all the involved RVs are Gaussian and conditionally independent, their sum results in a Gaussian RV with a mean and variance that is the sum of means and variances, respectively, and we obtain
\begin{align}
    \begin{split}
    &f_{ \mathtt{P}_{i}+ \mathtt{E}|s_{i-M+1}^{i-1}}(\omega) =  g\bigg(\omega;\mu_{\mathtt{E}}+N_{\mathrm{T}}\sum_{j=2}^{M}s_{i-j+1}h_{j},\\& \hspace{2.7cm}\sigma_{\mathtt{E}}^{2}+N_{\mathrm{T}}\sum^{M}_{j=2}s_{i-j+1}h_{j}(1-h_{j})\bigg)~,
    \end{split}
\end{align}
\begin{align}
    \begin{split}
    f_{\mathtt{C}_{i}+ \mathtt{P}_{i}+\mathtt{E}|s_{i-M+1}^{i}}(\omega) = & g\bigg(\omega;\mu_{\mathtt{E}}+N_{\mathrm{T}}\sum^{M}_{j=1}s_{i-j+1}h_{j},\\& \hspace{-0.3cm}\sigma_{\mathtt{E}}^{2}+N_{\mathrm{T}} 
    \sum^{M}_{j=1}s_{i-j+1}h_{j}(1-h_{j})\bigg)~.
    \end{split}
\end{align}
Thus, the channel transition probabilities given a specific sequence of symbols can be written as
\begin{align}
\begin{split}
    &P_{\hat{S}_{i}|S_{i-M+1}^{i-1},S_{i}}(1|s_{i-M+1}^{i-1},0)= \\&\hspace{0.7cm}\Pr(\mathtt{P}_{i}+\mathtt{E} \geq\tau |s_{i-M+1}^{i-1}) = \\& \hspace{1.4cm} Q\Bigg(\frac{\tau-\mu_{\mathtt{E}}-N_{\mathrm{T}}\sum^{M}_{j=2}s_{i-j+1}h_{j}}{\sqrt{\sigma_{\mathtt{E}}^{2}+N_{\mathrm{T}}\sum^{M}_{j=2}s_{i-j+1}h_{j}(1-h_{j})}}\Bigg)~,
\end{split}
\end{align}
\begin{align}
\begin{split}
    &P_{\hat{S}_{i}|S_{i-M+1}^{i-1},S_{i}}(0|s_{i-M+1}^{i-1},0)=\\&\hspace{1.5cm} \Pr(\mathtt{P}_{i}+\mathtt{E} <\tau |s_{i-M+1}^{i-1} ) = \\&\hspace{3cm} 1-\Pr(\mathtt{P}_{i}+\mathtt{E} \geq\tau |s_{i-M+1}^{i-1})~,
\end{split}
\end{align}
\begin{align}
\begin{split}
    & P_{\hat{S}_{i}|S_{i-M+1}^{i-1},S_{i}}(1|s_{i-M+1}^{i-1},1) =\\&\hspace{0.75cm} \Pr(\mathtt{C}_{i}+\mathtt{P}_{i}+\mathtt{E} \geq\tau |s_{i-M+1}^{i}) =\\&\hspace{1.5cm} Q\Bigg(\frac{\tau-\mu_{\mathtt{E}}-N_{\mathrm{T}}\sum^{M}_{j=1}s_{i-j+1}h_{j}}{\sqrt{\sigma_{\mathtt{E}}^{2}+N_{\mathrm{T}}\sum^{M}_{j=1}s_{i-j+1}h_{j}(1-h_{j})}}\Bigg)~,
\end{split}
\end{align}
\begin{align}
\begin{split}
    & P_{\hat{S}_{i}|S_{i-M+1}^{i-1},S_{i}}(0|s_{i-M+1}^{i-1},1) =\\&\hspace{1.3cm} \Pr(\mathtt{C}_{i}+\mathtt{P}_{i}+\mathtt{E} <\tau |s_{i-M+1}^{i}) =\\&\hspace{2.6cm} 1 - \Pr(\mathtt{C}_{i}+\mathtt{P}_{i}+\mathtt{E} \geq\tau |s_{i-M+1}^{i})~.
\end{split}
\end{align}
\section{Achievable Information Rate Analysis} \label{sec:capacity}
\par In this section, we define and derive the MI, AIR, and capacity for two different types of sources and two degrees of knowledge about the preceding transmitted symbols at the receiver side. The sources under investigation are a correlated source that generates symbols based on the first-order Markov process, and a source that generates symbols independently. For each source, we derive two MIs that correspond to different degrees of knowledge about the ISI at the receiver side. The first case, termed ISI-Aware scenario, assumes that the receiver knows the previously transmitted symbols within the memory intervals, which is equivalent to knowing the terms contributing to ISI. The second scenario assumes that the receiver is uncertain about the previously transmitted symbols, corresponding to not knowing the ISI terms. This is referred as ISI-Unaware scenario. Before delving into these four scenarios, we need to establish the formalism for calculating the MI, AIR, and channel capacity. Actually, a unique feature of the CIR introduced in Sec.~\ref{sec:system_model} and shown in Fig.~\ref{fig:CNAP}, was its variation based on $T_{\mathrm{sym}}$. Calculation of the AIR facilitates the exploration of optimal $T_{\mathrm{sym}}$ that allows us to transmit information at the highest possible rate per unit of time.
\par The channel capacity, $C$,  is defined as the maximization of the AIR over the input distributions, $P_{S}$: 
\begin{equation}
    C = \max_{P_{S}}\text{AIR}~.\label{eq:capacity}
\end{equation}
The AIR is defined as the MI between the input process $S$$\,=\,$$(S_{1},S_{2},\ldots)$ and output process $\hat{S}$$\,=\,$$(\hat{S}_{1},\hat{S}_2,\ldots)$ divided by the STI $T_{\mathrm{sym}}$. 
\begin{equation}
    \text{AIR}=\frac{I(S;\hat{S})}{T_{\mathrm{sym}}}~.
\end{equation}
The MI between the input process $S$$\,=\,$$(S_{1},S_{2},\ldots)$ and output process $\hat{S}$$\,=\,$$(\hat{S}_{1},\hat{S}_2,\ldots)$ of time-invariant channel with memory can be written as~\cite{arnold2003simulation}
\begin{align}
\begin{split}
    I(S;\hat{S}) &= \lim_{n\to\infty}\frac{1}{n}I(S_{1},\ldots,S_{n};\hat{S}_{1},\ldots,\hat{S}_{n})~,\\&
    = \lim_{n\to\infty}\frac{1}{n}I(S_{1}^{n};\hat{S}_{1}^{n})~.
    \label{eq:MIn}
    \end{split}
\end{align}
The expression given in~\eqref{eq:MIn} evaluates the average of the MI per channel use. In more detail, the MI is calculated between the input-output processes across an infinitely long symbols sequence and then divided by the length of the sequence. 
\par The MI can be written as the difference between the entropy of the input sequence and the entropy of the input given the output sequence
\begin{equation}
    I(S_{1}^{n};\hat{S}_{1}^{n}) = H(S_{1}^{n}) - H(S_{1}^{n}|\hat{S}_{1}^{n})~.
    \label{eq:Hs}
\end{equation}
From the chain rule of entropy, we can write~\eqref{eq:Hs} as
\begin{align}
    \begin{split}
    I(S_{1}^{n};\hat{S}_{1}^{n}) = \sum_{i=1}^{n} H(S_{i}|S_{1}^{i-1}) - H(S_{i}|S_{1}^{i-1},\hat{S}_{1}^{n})~.
    \end{split}
\end{align}
In conventional communication systems, information is typically transmitted in the form of a sequence, and the correlation between the received signal at a given time and the received signal observed before and after that time interval is utilized. Hence, there is a dual-side correlation exploitation regardless of time. However, in this study, we employ the causality assumption for detection in the communication system. Consequently, we discard the estimated samples following the $i$th interval (\textit{i.e.}, current time).
\begin{align}
    \begin{split}
    &\sum_{i=1}^{n} H(S_{i}|S_{1}^{i-1}) - H(S_{i}|S_{1}^{i-1},\hat{S}_{1}^{i})=\\& \hspace{0.6cm} \sum_{i=1}^{n} H(S_{i}|S_{1}^{i-1}) - H(S_{i}|S_{1}^{i-1},\hat{S}_{i})\leq I(S_{1}^{n};\hat{S}_{1}^{n})  ~.
    \end{split}
\end{align}
Once the previously transmitted symbols, $S_{1}^{i-1}$, are known, there is no longer any uncertainty in the previously detected symbols, $\hat{S}_{1}^{i-1}$. Consequently, we can safely discard the previously detected symbols as they contain redundant information.
\subsection{Correlated Source}
One of the primary obstacles in communication systems involving memory lies in analyzing the impact of bursty symbol transmission. To delve into the MI of the communication system when symbols are transmitted in a bursty fashion, we consider a source with memory that generates symbols while taking into account temporal correlations between them. In this study, we employ a first-order Markov process to model the correlated source, as illustrated in the following representation
\begin{center}
	\begin{tikzpicture}[->, >=stealth', auto, semithick, node distance=3cm]
	\tikzstyle{every state}=[fill=white,draw=black,thick,text=black,scale=1]
	\node[state]    (A)               {$0$};
	\node[state]    (B)[right of=A]   {$1$};
	\path
	(A) edge[loop left]			node{$1-p$}	(A)
        edge[bend left,above]	node{$p$}	(B)
	(B) edge[bend left,below]	node{$q$}	(A)
	edge[loop right]		    node{$1-q$}	(B);
	\end{tikzpicture}
\end{center}
where the transition probabilities are $1-p$$\,=\,$$P_{S_{i}|S_{i-1}}(0|0)$, $p$$\,=\,$$P_{S_{i}|S_{i-1}}(1|0)$, $q$$\,=\,$$P_{S_{i}|S_{i-1}}(0|1)$, and $1-q$$\,=\,$$P_{S_{i}|S_{i-1}}(1|1)$. Mathematically, the temporal evolution of the Markov source shown in the preceding representation can be expressed as
 \begin{equation}
      \begin{bmatrix}
        P_{S_{i}}(0)\\ 
        P_{S_{i}}(1)
    \end{bmatrix}
    =
    \begin{bmatrix}
        1-p & q\\ 
        p & 1-q
    \end{bmatrix}
    \begin{bmatrix}
        P_{S_{i-1}}(0) \\
        P_{S_{i-1}}(1)
    \end{bmatrix}.\label{eq:MKVmat}
 \end{equation}
Accordingly, one can compute the asymptotical/stationary probability of the two symbols (\textit{i.e.} the probability of $``1"$s and $``0"$s) generated by the source in a sufficiently long sequence as
\begin{equation}
    \pi_{0} = \frac{q}{p+q},\label{eq:pi0stationany}
\end{equation}
\begin{equation}
    \pi_{1} = \frac{p}{p+q}.\label{eq:pi1stationany}
\end{equation}
In the case of the first-order Markov source, we can simplify the sum of the entropy of the sequence generated by the source as
\begin{align}
\begin{split}
    \sum_{i=1}^{n} H(S_{i}|S_{1}^{i-1}) &= H(S_{1}) + \sum_{i=2}^{n} H(S_{i}|S_{1}^{i-1})~,\\&
    =H(S_{1}) +(n-1) H(S_{i}|S_{1}^{i-1})~,\\&
    =H(S_{1}) +(n-1) H(S_{i}|S_{i-1})~.
    \label{eq:sourceH}
\end{split}
\end{align}
Substituting~\eqref{eq:sourceH} into the definition of entropy of the process~\eqref{eq:MIn} and taking the limit, we can write the average entropy of the source as
\begin{align}
\begin{split}
    &\hspace{-0.28cm}\lim_{n\to\infty}\frac{1}{n}\sum_{i=1}^{n} H(S_{i}|S_{1}^{i-1}) =\\&\hspace{-0.28cm}\lim_{n\to\infty}\frac{1}{n} \Big(H(S_{1})+(n-1) H(S_{i}|S_{i-1})\Big)
    = H(S_{i}|S_{i-1})~.\label{eq:HCoSrc}
\end{split}
\end{align}
In this study, we consistently make the assumption that the initial symbol in a sequence generated by the Markov source follows the asymptotic probabilities of the source. Consequently, we express the entropy of the Markov source as the mean of the two conditional entropies.
\begin{align}
\begin{split}
    H(S_{i}|S_{i-1}) =& P_{S_{i-1}}(0)H(S_{i}|S_{i-1}=0)+\\&\hspace{-1.5cm} P_{S_{i-1}}(1)H(S_{i}|S_{i-1}=1)
    =\pi_{0}H_{2}(p)+\pi_{1}H_{2}(q)~.
\end{split}
\end{align}
\subsubsection{ISI-Aware}
In this scenario, we consider the presence of a receiver which has a comprehensive knowledge about the previously transmitted symbols. This knowledge is tantamount to being aware of the factors responsible for ISI. Analogous to the source entropy, we can express the sum of conditional entropy of the transmitted symbol given the previously transmitted sequence and the currently estimated symbol as the sum of individual entropies.
\begin{align}
    \begin{split}
        &\sum_{i=1}^{n} H(S_{i}|S_{1}^{i-1},\hat{S}_{i}) = \\& H(S_{1}|\hat{S}_{1})+H(S_{2}|S_{1}^{2},\hat{S}_{2}) + \cdots+\sum_{i=M}^{n} H(S_{i}|S_{1}^{i-1},\hat{S}_{i}) 
    =\\& H(S_{1}|\hat{S}_{1})+H(S_{2}|S_{1}^{2},\hat{S}_{2}) + \cdots+(n-M)H(S_{i}|S_{1}^{i-1},\hat{S}_{i})~.
    \end{split}
\end{align}
Applying the limit to take the average results in
\begin{align}
\begin{split}
    &\lim_{n\to\infty}\frac{1}{n}\sum_{i=1}^{n} H(S_{i}|S_{1}^{i-1},\hat{S}_{i}) =\\&\hspace{0.5cm} \lim_{n\to\infty}\frac{1}{n} \Big(H(S_{1}|\hat{S}_{1})+H(S_{2}|S_{1}^{2},\hat{S}_{2}) + \cdots\\&\hspace{0.7cm}+(n-M)H(S_{i}|S_{1}^{i-1},\hat{S}_{i})\Big)=
    H(S_{i}|S_{1}^{i-1},\hat{S}_{i})~.
\end{split}
\end{align}
Considering the finite length of the memory interval as previously defined in Sec.~\ref{sec:system_model}, we can disregard the level of surprise associated with symbols transmitted significantly earlier. Consequently, we discard the symbols transmitted prior to the effective memory interval.
\begin{equation}
    H(S_{i}|S_{1}^{i-1},\hat{S}_{i}) \leq H(S_{i}|S_{i-M+1}^{i-1},\hat{S}_{i})~.
\end{equation}
The entropy of the current symbol, conditioned on the previously transmitted symbols and the currently estimated symbol, can be expressed as the marginalization over the realizations of the previously transmitted symbols and the currently estimated symbol, yielding
\begin{align}
    \begin{split}
    &H(S_{i}|S_{i-M+1}^{i-1},\hat{S}_{i}) =\\& \sum_{\forall s_{i-M+1}^{i-1},\hat{s}_{i}}P_{S_{i-M+1}^{i-1},\hat{S}_{i}}(s_{i-M+1}^{i-1},\hat{s}_{i})H(S_{i}|s_{i-M+1}^{i-1},\hat{s}_{i})~,\label{eq:SGpreShatS}
\end{split}
\end{align}
and
\begin{align}
    \begin{split}
    &H(S_{i}|s_{i-M+1}^{i-1},\hat{s}_{i}) =\\&\hspace{2.3cm} -\sum_{\forall s_{i}}P_{S_{i}|S_{i-M+1}^{i-1},\hat{S}_{i}}(s_{i}|s_{i-M+1}^{i-1},\hat{s}_{i})\times \\&\hspace{2.3cm}  \log_{2}\big(P_{S_{i}|S_{i-M+1}^{i-1},\hat{S}_{i}}(s_{i}|s_{i-M+1}^{i-1},\hat{s}_{i})\big)~.\label{eq:PA|BC}
\end{split}
\end{align}
The conditional probability of realizations in~\eqref{eq:PA|BC} can be computed as follows (see Appendix~\ref{FirstAppendix})
\begin{align}
    \begin{split}
    &P_{S_{i}|S_{i-M+1}^{i-1},\hat{S}_{i}}(s_{i}|s_{i-M+1}^{i-1},\hat{s}_{i}) = \\& \frac{P_{\hat{S}_{i}|S_{i-M+1}^{i}}(\hat{s}_{i}|s_{i-M+1}^{i})P_{S_{i}|S_{i-1}}(s_{i}|s_{i-1})}{\sum\limits_{\substack{x\in\{0,1\}}}\hspace{-.1cm}P_{\hat{S}_{i}|S_{i-M+1}^{i-1},S_{i}}(\hat{s}_{i}|s_{i-M+1}^{i-1},x)P_{S_{i}|S_{i-1}}(x|s_{i-1})}~,
\end{split}
\end{align}
where the conditional and joint probabilities of a specific estimated symbol and previously transmitted symbols are
\begin{align}
    \begin{split}
    &P_{\hat{S}_i|S_{i-M+1}^{i-1}}(\hat{s}_i|s_{i-M+1}^{i-1})=\\&\hspace{0.3cm} \sum\limits_{\substack{\forall s_{i}}}P_{\hat{S}_{i}|S_{i-M+1}^{i-1},S_{i}}(\hat{s}_{i}|s_{i-M+1}^{i-1},s_{i})P_{S_{i}|S_{i-1}}(s_{i}|s_{i-1})~,
\end{split}
\end{align}
\begin{align}
    \begin{split}
    &P_{\hat{S}_i,S_{i-M+1}^{i-1}}(\hat{s}_i,s_{i-M+1}^{i-1}) =\\&\hspace{1.6cm} P_{\hat{S}_i|S_{i-M+1}^{i-1}}(\hat{s}|s_{i-M+1}^{i-1})P_{S_{i-M+1}^{i-1}}(s_{i-M+1}^{i-1})~.\label{eq:hatSJpreS}
\end{split}
\end{align}
Based on~\eqref{eq:hatSJpreS}, the calculation of the probability for a given symbol sequence generated by the source requires knowledge of the Markov model. Since we are utilizing a Markov source as described in~\eqref{eq:MKVmat}, the probability of a specific sequence can be determined by traversing the sequence through the Markov model. It is important to note that the probability of the first element in a sequence is assumed to correspond to the asymptotic probability of the Markov source for that particular symbol realization:
\begin{equation}
    P_{S_{r}^{v}}(s_{r}^{v}) = 
    \pi_{0}^{1-s_{r}} \pi_{1}^{s_{r}} \prod\limits_{\substack{j=r+1}}^{v} P_{S_{j}|S_{j-1}}(s_{j}|s_{j-1})~.
\end{equation}
In the end, the MI associated with the correlated source and ISI awareness is~\eqref{eq:I_ISIA_CRR}.
%
Note that the MI cannot have a negative value, and what we are computing in this paper is equivalent to lower bounds to the actual MIs due to the assumptions such as causality, effective memory, etc. Hence, we only take into account the positive values of MIs.

\begin{figure*}[t!]
\begin{align}
\begin{split}&I_{\text{ISIA}}^{\text{CRR}} = \Bigg\{\pi_{0}H_{2}(p)+\pi_{1}H_{2}(q) +
    \sum_{\forall s_{i-M+1}^{i-1},\hat{s}_{i}}\Bigg[\sum\limits_{\substack{\forall s_{i}}}\bigg[P_{\hat{S}_{i}|S_{i-M+1}^{i-1},S_{i}}(\hat{s}_{i}|s_{i-M+1}^{i-1},s_{i})P_{S_{i}|S_{i-1}}(s_{i}|s_{i-1})\bigg]
    \pi_{0}^{1-s_{i-M+1}} \times\\&\hspace{2cm} \pi_{1}^{s_{i-M+1}} \prod\limits_{\substack{j=i-M+2}}^{i-1} \Big[P_{S_{j}|S_{j-1}}(s_{j}|s_{j-1})\Big]
    \sum_{\forall s_{i}}\bigg[\frac{P_{\hat{S}_{i}|S_{i-M+1}^{i-1},S_{i}}(\hat{s}_{i}|s_{i-M+1}^{i-1},s_{i})P_{S_{i}|S_{i-1}}(s_{i}|s_{i-1})}{\sum\limits_{\substack{x\in\{0,1\}}}P_{\hat{S}_{i}|S_{i-M+1}^{i-1},S_{i}}(\hat{s}_{i}|s_{i-M+1}^{i-1},x)P_{S_{i}|S_{i-1}}(x|s_{i-1})}\times\\&\hspace{7cm}
    \log_{2}\Big(\frac{P_{\hat{S}_{i}|S_{i-M+1}^{i-1},S_{i}}(\hat{s}_{i}|s_{i-M+1}^{i-1},s_{i})P_{S_{i}|S_{i-1}}(s_{i}|s_{i-1})}{\sum\limits_{\substack{x\in\{0,1\}}}P_{\hat{S}_{i}|S_{i-M+1}^{i-1},S_{i}}(\hat{s}_{i}|s_{i-M+1}^{i-1},x)P_{S_{i}|S_{i-1}}(x|s_{i-1})}\Big)\bigg]\Bigg]\Bigg\}^{+}~.\label{eq:I_ISIA_CRR}
\end{split}
\end{align}
\hrule
\end{figure*}
\subsubsection{ISI-Unaware}\label{sec:2.1.2}
In this particular scenario, we make the assumption that the receiver does not have any knowledge regarding the symbols transmitted prior to the current time. This assumption is equivalent to loosening the bound on MI. Consequently, by disregarding the information pertaining to previously transmitted symbols, we can establish the following inequality
\begin{equation}
    H(S_{i}|S_{i-M+1}^{i-1},\hat{S}_{i})\leq  H(S_{i}|\hat{S}_{i})~,\label{eq:ISIA<ISIU}
\end{equation}
and from the definition of average conditional entropy, we write
\begin{equation}
    H(S_{i}|\hat{S}_{i}) = -\sum_{\forall s_{i},\hat{s}_{i}}P_{S_{i},\hat{S}_{i}}({s_{i},\hat{s}_{i}})\log_{2}\big(P_{S_{i}|\hat{S}_{i}}({s_{i}|\hat{s}_{i}})\big)~,
\end{equation}
\begin{equation}
    P_{S_{i},\hat{S}_{i}}(s_{i},\hat{s}_{i}) = P_{\hat{S}_{i}|S_{i}}(\hat{s}_{i}|s_{i})P_{S_{i}}(s_{i})~.\label{eq:SiJhatS}
\end{equation}
The conditional probability of the detected symbol given the transmitted symbol is obtained by marginalizing over the previously transmitted symbols using the  Bayes theorem. 
\begin{align}
    \begin{split}
    &P_{\hat{S}_{i}|S_{i}}(\hat{s}_{i}|s_{i}) =\\& \sum_{\forall s^{i-1}_{i-M+1}}P_{\hat{S}_{i}|S^{i-1}_{i-M+1},S_{i}}(\hat{s}_{i}|s^{i-1}_{i-M+1},s_{i})\frac{P_{S_{i-M+1}^{i}}(s_{i-M+1}^{i})}{P_{S_{i}}(s_{i})}~.\label{eq:hatSGSi}
\end{split}
\end{align}
Substituting~\eqref{eq:hatSGSi} into~\eqref{eq:SiJhatS}, the term corresponding to the currently transmitted symbol cancel out, and we obtain
\begin{align}
    \begin{split}
    &P_{S_{i},\hat{S}_{i}}(s_{i},\hat{s}_{i}) =\\& \sum_{\forall s^{i-1}_{i-M+1}}P_{\hat{S}_{i}|S^{i-1}_{i-M+1},S_{i}}(\hat{s}_{i}|s^{i-1}_{i-M+1},s_{i})P_{S_{i-M+1}^{i}}(s_{i-M+1}^{i})~.
\end{split}
\end{align}
To compute the conditional probability of the current transmitted symbol given the estimated one, we apply the Bayes rule 
\begin{equation}
    P_{S_{i}|\hat{S}_{i}}({s_{i}|\hat{s}_{i}}) = \frac{P_{S_{i},\hat{S}_{i}}(s_{i},\hat{s}_{i})}{P_{\hat{S}_{i}}(\hat{s}_{i})}~.
\end{equation}
The probability of the estimated symbol can be computed from the marginalization of the joint probability over all possible realizations of the transmitted symbols.
\begin{equation}
    P_{\hat{S}_{i}}(s_{i}) = \sum_{\forall s_{i}}P_{\hat{S}_{i},S_{i}}(\hat{s}_{i},s_{i})~.
\end{equation}
In the end, one can compute the MI corresponding to the correlated source with ISI unawareness on the receiver side as~\eqref{eq:I_ISIU_CRR}.
\begin{figure*}[t!]
\begin{align}
\begin{split}
    &I_{\text{ISIU}}^{\text{CRR}} =\Bigg\{ \pi_{0}H_{2}(p)+\pi_{1}H_{2}(q) +
    \sum_{\forall s_{i},\hat{s}_{i}}\Bigg[\sum_{\forall s^{i-1}_{i-M+1}}\hspace{-0.3cm}\bigg[P_{\hat{S}_{i}|S^{i-1}_{i-M+1},S_{i}}(\hat{s}_{i}|s^{i-1}_{i-M+1},s_{i})\pi_{0}^{1-s_{i-M+1}} \pi_{1}^{s_{i-M+1}}
    \hspace{-0.4cm}\prod\limits_{\substack{j=i-M+2}}^{i-1} \hspace{-0.4cm}P_{S_{j}|S_{j-1}}(s_{j}|s_{j-1})\bigg]\times\\&
    \log_{2}\Big(\frac{\sum\limits_{\substack{\forall s^{i-1}_{i-M+1}}}\Big[P_{\hat{S}_{i}|S^{i-1}_{i-M+1},S_{i}}(\hat{s}_{i}|s^{i-1}_{i-M+1},s_{i})\pi_{0}^{1-s_{i-M+1}} \pi_{1}^{s_{i-M+1}}\prod\limits_{\substack{j=i-M+2}}^{i} P_{S_{j}|S_{j-1}}(s_{j}|s_{j-1})\Big]}{\sum\limits_{\substack{x\in\{0,1\}}}\Big[\sum\limits_{\substack{\forall s^{i-1}_{i-M+1}}}P_{\hat{S}_{i}|S^{i-1}_{i-M+1},S_{i}}(\hat{s}_{i}|s^{i-1}_{i-M+1},x)P_{S_{i}|S_{i-1}}(x|s_{i-1})\pi_{0}^{1-s_{i-M+1}} \pi_{1}^{s_{i-M+1}}\prod\limits_{\substack{j=i-M+2}}^{i-1} P_{S_{j}|S_{j-1}}(s_{j}|s_{j-1})\Big]}\Big)\Bigg]\Bigg\}^{+}~.\label{eq:I_ISIU_CRR}
\end{split}
\end{align}
\hrule
\end{figure*}
\subsection{Independent Source}
Another type of source that is considered in this paper from a statistical perspective is the one where symbols are generated independently with specific probabilities. Let $\lambda_{1}$$\,=\,$$P_{S_i}(1)$ and $\lambda_{0}$$\,=\,$$1-\lambda_{1}=P_{S_i}(0)$ denote the probabilities of transmitting symbols $``1"$ and $``0"$, respectively. In this scenario, there is no temporal dependency between the symbols generated by the source, \textit{i.e.},
\begin{equation}
    P_{S_1^n}(s_1^n) = \prod_{i=1}^n P_{S_i}(s_i).
\end{equation}
As a result, we can discard the conditioning on the previously transmitted symbol, and the entropy of the source simplifies to a binary entropy function.
\begin{equation}
    H(S_{i}|S_{1}^{i-1}) = H(S_{i}) = H_{2}(\lambda_{0})~.
\end{equation}
\subsubsection{ISI-Aware}
Similarly to the scenario with the correlated source, we can rely on Eqns.~\eqref{eq:SGpreShatS}-\eqref{eq:hatSJpreS}. Noting that the source is independent, a main difference in this case in comparison to the correlated source scenario is the probability of a particular sequence of symbols, and we write it as
\begin{equation}
    P_{S_{r}^{v}}(s_{r}^{v}) = \lambda_{0}^{v-r+1-w_{H}(S_{r}^{v})}\lambda_{1}^{w_{H}(S_{r}^{v})}~.\label{eq:indSrc}
\end{equation}
The MI for the case of an independent source with knowledge about the previously transmitted symbols (\textit{i.e.}, ISI-Aware) is~\eqref{eq:I_ISIA_IND}.
\begin{figure*}[b!]
\hrule
\begin{align}
\begin{split}
    I_{\text{ISIA}}^{\text{IND}} =& \Bigg\{H_{2}(\lambda_{0}) +
    \sum_{\forall s_{i-M+1}^{i-1},\hat{s}_{i}}\Bigg[\lambda_{0}^{M-1-w_{H}(s_{i-M+1}^{i-1})}\lambda_{1}^{w_{H}(s_{i-M+1}^{i-1})}\sum\limits_{\substack{\forall s_{i}}}\bigg[\lambda_{0}^{1-s_{i}} \lambda_{1}^{s_{i}}
    P_{\hat{S}_{i}|S_{i-M+1}^{i-1},S_{i}}(\hat{s}_{i}|s_{i-M+1}^{i-1},s_{i})\bigg]\times
    \\&
    \sum_{\forall s_{i}}\bigg[\frac{\lambda_{0}^{1-s_{i}} \lambda_{1}^{s_{i}}P_{\hat{S}_{i}|S_{i-M+1}^{i-1},S_{i}}(\hat{s}_{i}|s_{i-M+1}^{i-1},s_{i})}{\sum\limits_{\substack{x\in\{0,1\}}}\lambda_{0}^{1-x} \lambda_{1}^{x} P_{\hat{S}_{i}|S_{i-M+1}^{i-1},S_{i}}(\hat{s}_{i}|s_{i-M+1}^{i-1},x)}
    \log_{2}\Big(\frac{\lambda_{0}^{1-s_{i}} \lambda_{1}^{s_{i}}P_{\hat{S}_{i}|S_{i-M+1}^{i-1},S_{i}}(\hat{s}_{i}|s_{i-M+1}^{i-1},s_{i})}{\sum\limits_{\substack{x\in\{0,1\}}}\lambda_{0}^{1-x} \lambda_{1}^{x}P_{\hat{S}_{i}|S_{i-M+1}^{i-1},S_{i}}(\hat{s}_{i}|s_{i-M+1}^{i-1},x)}\Big)\bigg]\Bigg]\Bigg\}^{+}~.\label{eq:I_ISIA_IND}
\end{split}
\end{align}    
\end{figure*}
\subsubsection{ISI-Unaware}
Without the knowledge of previously transmitted symbols, the equations derived in Sec.~\ref{sec:2.1.2} remain applicable. Nevertheless, it is necessary to calculate the probability of each specific sequence using~\eqref{eq:indSrc}, considering the independent nature of symbol generation by the source. Consequently, the MI of the independent source, under the condition of unknown previously transmitted symbols, can be extracted~\eqref{eq:I_ISIU_IND}.
\begin{figure*}[t!]
\begin{align}
\begin{split}
    I_{\text{ISIU}}^{\text{IND}} =& \Bigg\{H_{2}(\lambda_{0})+
    \sum\limits_{\substack{\forall s_{i},\hat{s}_{i}}}\Bigg[\sum\limits_{\substack{\forall s^{i-1}_{i-M+1}}}\bigg[\lambda_{0}^{M-1-w_{H}(s_{i-M+1}^{i-1})}\lambda_{1}^{w_{H}(s_{i-M+1}^{i-1})}P_{\hat{S}_{i}|S^{i-1}_{i-M+1},S_{i}}(\hat{s}_{i}|s^{i-1}_{i-M+1},s_{i})\bigg]\times
    \\&\hspace{2cm}
    \log_{2}\Big(\frac{\sum\limits_{\substack{\forall s^{i-1}_{i-M+1}}}\lambda_{0}^{M-w_{H}(s_{i-M+1}^{i-1})-s_{i}}\lambda_{1}^{w_{H}(s_{i-M+1}^{i-1})+s_{i}}P_{\hat{S}_{i}|S^{i-1}_{i-M+1},S_{i}}(\hat{s}_{i}|s^{i-1}_{i-M+1},s_{i})}{\sum\limits_{\substack{x\in\{0,1\}}}\sum\limits_{\substack{\forall s^{i-1}_{i-M+1}}}\lambda_{0}^{M-w_{H}(s_{i-M+1}^{i-1})-x}\lambda_{1}^{w_{H}(s_{i-M+1}^{i-1})+x}P_{\hat{S}_{i}|S^{i-1}_{i-M+1},S_{i}}(\hat{s}_{i}|s^{i-1}_{i-M+1},s_{i})}\Big)\Bigg]\Bigg\}^{+}~.\label{eq:I_ISIU_IND}
\end{split}
\end{align}
\hrule
\end{figure*}
\section{Numerical Evaluation and Results}\label{sec:simulation_results}
We present a selection of results that illustrate the superiority of correlated sources in achieving higher capacity. It should be noted that the optimal input distribution for achieving capacity may not be uniform. The numerical evaluation was conducted using system parameters listed in Table~\ref{tab:param}, obtained from~\cite{vakilipoor2022asymptotic}, with the exception of the external noise and $\alpha$. We intentionally selected noise standard deviation $\sigma_{\text{ext}}$ and mean $\mu_{\text{ext}}$ such that there are instances where the values of $\mathtt{E}$ become negative, indicating that the external noise impedes IPs absorption. The parameter $\alpha$ is chosen to ensure the validity of the last sample of the CIR as~\eqref{eq:valid_apx}.
\begin{table}[!t]
\begin{center}
\caption{System parameters}
\label{tab:param}
\resizebox{1\columnwidth}{!}{
 \begin{tabular}{|| c | c | c ||}
 \hline
 Variable & Definition & Value \\ [0.5ex] 
 \hline\hline
  $N_{\mathrm{T}}$ & Number of released molecules & $10^{4}$ \\ 
  \hline
 $R$ & Radius of the receiver $\mathcal{R}_i$ & $1$ $\mu$m\\ 
 \hline
 $d$ & Distance between transmitter and center of receiver & $10$ $\mu$m \\
 \hline
 $\alpha$ & Minimum acceptable probability & $0.001$ \\
 \hline
   $\mu_{\text{ext}}$ & Mean of the external noise signal & $50$ \\
 \hline
   $\sigma_{\text{ext}}$ & Standard deviation of the external noise signal & $50$ \\
 \hline
 $D$ & Diffusion coefficient for the signaling molecule & $79.4$ $\mu\text{m}^2/\text{s}$ \\
 \hline
\end{tabular}}
\end{center}
\end{table}
\begin{figure}
    \centering
    \includegraphics[width=1\linewidth]{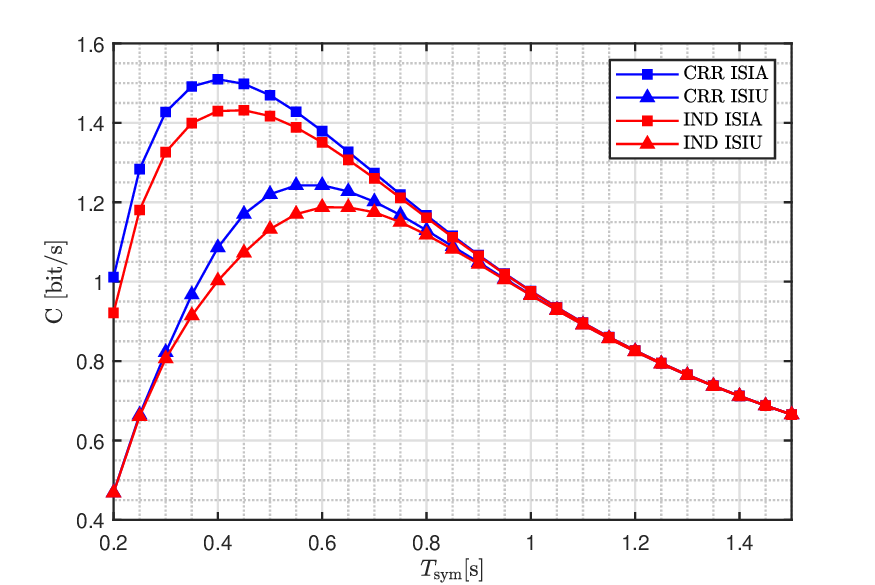}
    \caption{ Capacity, $C$, as a function of STI for four different scenarios as ISI-Awaren with correlated source, $\text{CRR~ISIA}$, (blue curve with square marker), ISI-Unaware with correlated source, $\text{CRR~ISIU}$, (blue curve with triangle marker), ISI-Aware with independent source, $\text{IND~ISIA}$, (red curve with square marker), and ISI-Unaware with independent source, $\text{IND~ISIU}$, (red curve with triangle marker).}
    \label{fig:Capacity}
\end{figure}
\par Figure~\ref{fig:Capacity} illustrates the channel capacity~\eqref{eq:capacity} for various STIs ($0.2$$\,\leq\,$$T_{\mathrm{sym}}$$\,\leq\,$$ 1.5$) in different scenarios. These scenarios include ISI-Aware with a correlated source (blue curve with square marker), ISI-Unaware with a correlated source (blue curve with triangle marker), ISI-Aware with an independent source (red curve with square marker), and ISI-Unaware with independent source (red curve with triangle marker).

As expected, the capacity with ISI awareness is generally higher than that with ISI unawareness. Interestingly, the correlated source achieves a higher capacity compared to the independent source. Normally, it is expected that hiring an independent source results in higher capacities in communication systems. However, in this unique scenario, due to the high ISI effect, the correlated source allows us to tackle the problem of the ISI, and the reduction of the source entropy compared to the independent one is worth it. Specifically, the maximum channel capacity in ISI-Aware scenario with correlated source, $C_{\text{ISIA}}^{\text{CRR}}$, is $1.50\,\text{[bit/s]}$ at $T_{\mathrm{sym}}$$\,=\,$$0.40$\,s, with the input probability distribution of $p$$\,=\,$$0.60$ and $q$$\,=\,$$0.62$. On the other hand, the maximum capacity for the ISI-Aware scenario with independent source, $C_{\text{ISIA}}^{\text{IND}}$, is $1.43\,\text{[bit/s]}$ at $T_{\mathrm{sym}}$$\,=\,$$0.45$\,s, with an input probability distribution of $\lambda_{0}$$\,=\,$$0.52$.

Comparing the two maximum capacities in the ISI-Aware scenario, we observe that the independent source achieves its maximum at a higher $T_{\mathrm{sym}}$ compared to the correlated source. 

Moving on to the ISI-Unaware scenario, the maximum capacity for the correlated source, $C_{\text{ISIU}}^{\text{CRR}}$, is $1.24\,\text{[bit/s]}$ at $T_{\mathrm{sym}}$$\,=\,$$0.57$\,s, with an optimum input probability distribution at $p$$\,=\,$$0.60$ and $q$$\,=\,$$0.60$. The maximum capacity for the independent source, $C_{\text{ISIA}}^{\text{IND}}$, is $1.18\,\text{[bit/s]}$ at $T_{\mathrm{sym}}$$\,=\,$$0.60$\,s, with an input probability distribution of $\lambda_{0}$$\,=\,$$0.50$. We also observe a slight shift in the STI corresponding to the maximum capacity in both ISI-Unaware cases.

Interestingly, by increasing the STI and consequently reducing the impact of ISI, all capacities overlap. Therefore, regardless of the source type or the ISI knowledge, the same performance can be achieved. This overlap occurs because as $T_{\mathrm{sym}}$ increases, the effect of ISI diminishes, rendering the knowledge of previously transmitted symbols less valuable. It is important to note that the significant difference between the ISI-Aware cases corresponding to different types of receivers is observed only within a certain range of STIs ($0.3$ to $0.85$). The same observation applies to the other two curves representing ISI unawareness. 
\begin{figure}
    \centering
    \includegraphics[width=1\linewidth]{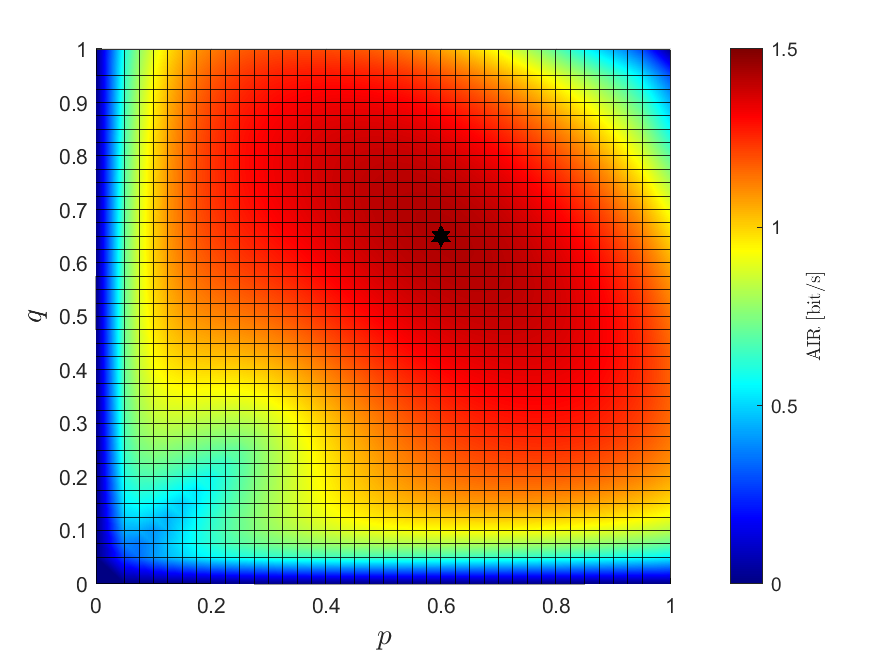}
    \caption{AIR as a function of the correlated source transition probabilities, $p$ and $q$, with~$T_{\mathrm{sym}}$$\,=\,$$0.3$\,s in ISI-Aware scenario. Capacity is obtained at $p$$\,=\,$$0.60$ and $q$$\,=\,$$0.65$, and it is $C_{\text{ISIA}}^{\text{CRR}}$$\,=\,$$1.42$.}
    \label{fig:C_ISIA_CRR_Tsym3e-1}
\end{figure}
\par Figure~\ref{fig:C_ISIA_CRR_Tsym3e-1} illustrates the AIR across the input distribution space, represented by color, in the context of ISI awareness with a correlated source. The black hexagram marker indicates the capacity point associated with this scenario, corresponding to $p$$\,=\,$$0.60$ and $q$$\,=\,$$0.65$, with a value of $C_{\text{ISIA}}^{\text{CRR}}$$\,=\,$$1.42$.

By analyzing the expressions for stationary probabilities~\eqref{eq:pi0stationany} and~\eqref{eq:pi1stationany}, we can infer that when $p$ and $q$ are equal to each other, it is equivalent to transmit an infinitely long sequence with equiprobable symbols. However, as $p$ and $q$ approach $1$, it indicates a preference to avoid consecutive transmission of the same symbol. In the provided example, we observe that the optimal input distribution is when the input distribution of the correlated source is asymptotically equiprobable, but it is preferable to avoid generating consecutive identical symbols, particularly for the transmission of $``1"$. This observation is supported by the fact that $q$ is slightly higher than $p$, indicating a lesser desire for transmitting two successive $``1"$ symbols. Of course, when both $p$ and $q$ are close to $1$, little information is transmitted since, after transmitting a $``1"$, a $``0"$ follows with high probability, and vice versa.
\begin{figure}
    \centering
    \includegraphics[width=1\linewidth]{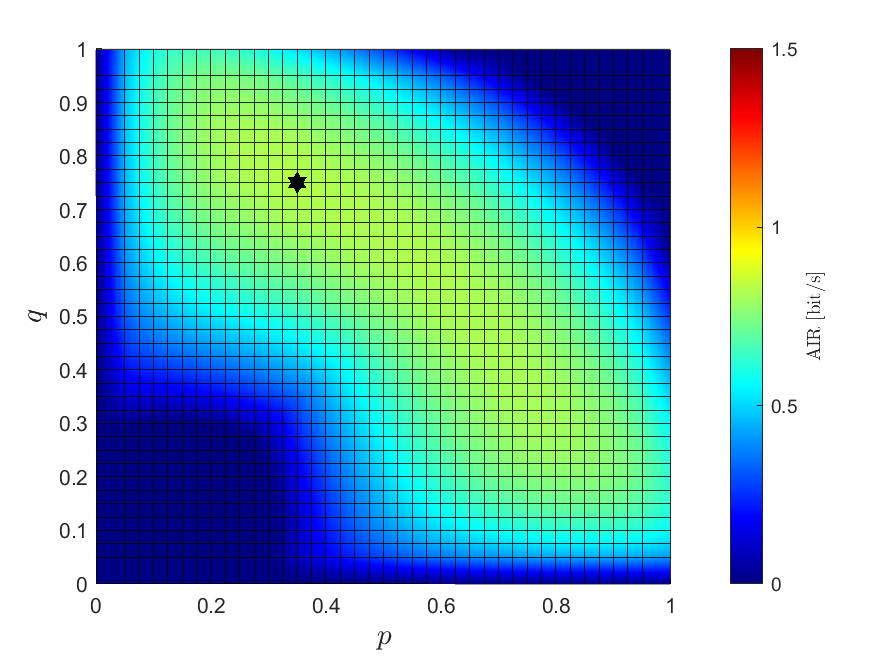}
    \caption{AIR as a function of the correlated source transition probabilities with $T_{\mathrm{sym}}$$\,=\,$$0.3$\,s in ISI-Unaware scenario. Capacity is achieved at $p$$\,=\,$$0.35$ and $q$$\,=\,$$0.75$, with a value of $C_{\text{ISIU}}^{\text{CRR}}$$\,=\,$$0.82$.}
    \label{fig:C_ISIU_CRR_Tsym3e-1}
\end{figure}
\par
\begin{figure}
    \centering
    \includegraphics[width=1\linewidth]{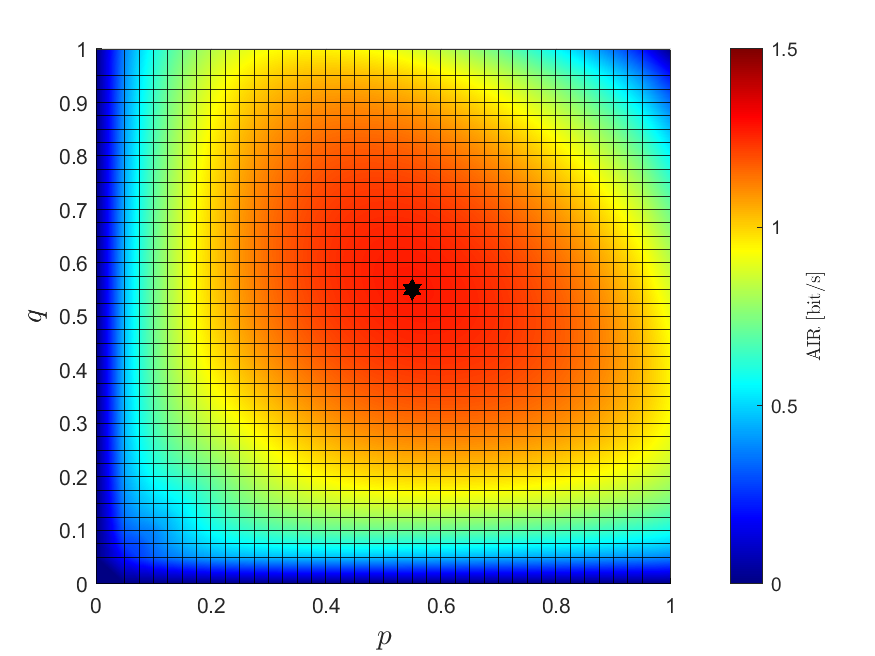}
    \caption{AIR as a function of the correlated source input distribution with $T_{\mathrm{sym}}$$\,=\,$$0.7$\,s in ISI-Aware scenario. Capacity is achieved at $p$$\,=\,$$q$$\,=\,$$0.55$, with a value of $C_{\text{ISIA}}^{\text{CRR}}$$\,=\,$$1.27$.}
    \label{fig:C_ISIA_CRR_Tsym7e-1}
\end{figure}

\begin{figure}
    \centering
    \includegraphics[width=1\linewidth]{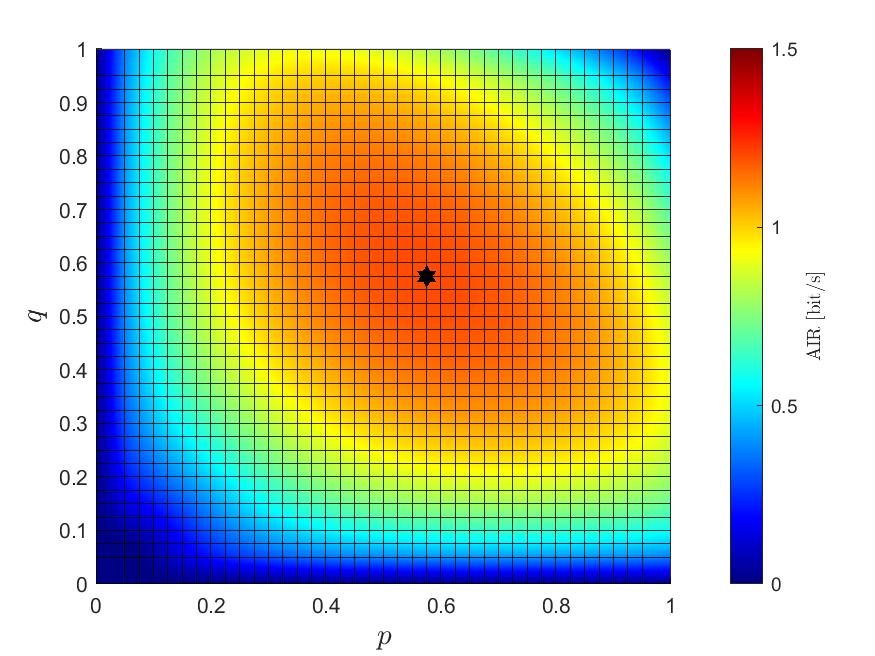}
    \caption{AIR as a function of the correlated source transition probabilities with $T_{\mathrm{sym}}$$\,=\,$$0.7$\,s in ISI-Unaware scenario. Capacity is $C_{\text{ISIU}}^{\text{CRR}}$ obtained with $p$$\,=\,$$0.57$ and $q$$\,=\,$$0.57$.}
    \label{fig:C_ISIU_CRR_Tsym7e-1}
\end{figure}

\begin{figure}
    \centering
    \includegraphics[width=1\linewidth]{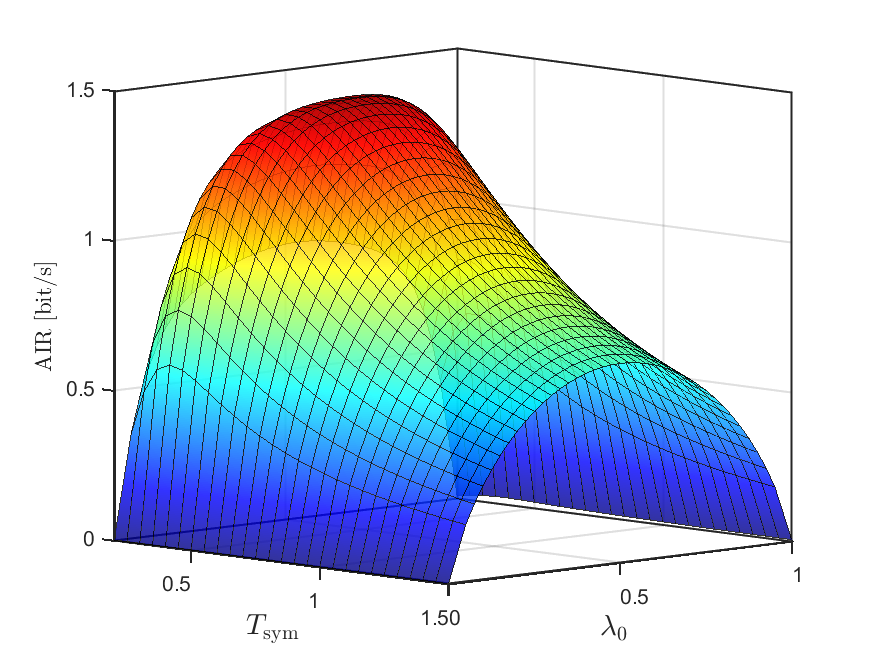}
    \caption{AIR as a function of the independent source probability of transmitting~$``0"$, $\lambda_0$, and of the STI $T_{\mathrm{sym}}$ in ISI-Aware scenario.}
    \label{fig:IND_AWARE}
\end{figure}

\begin{figure}
    \centering
    \includegraphics[width=1\linewidth]{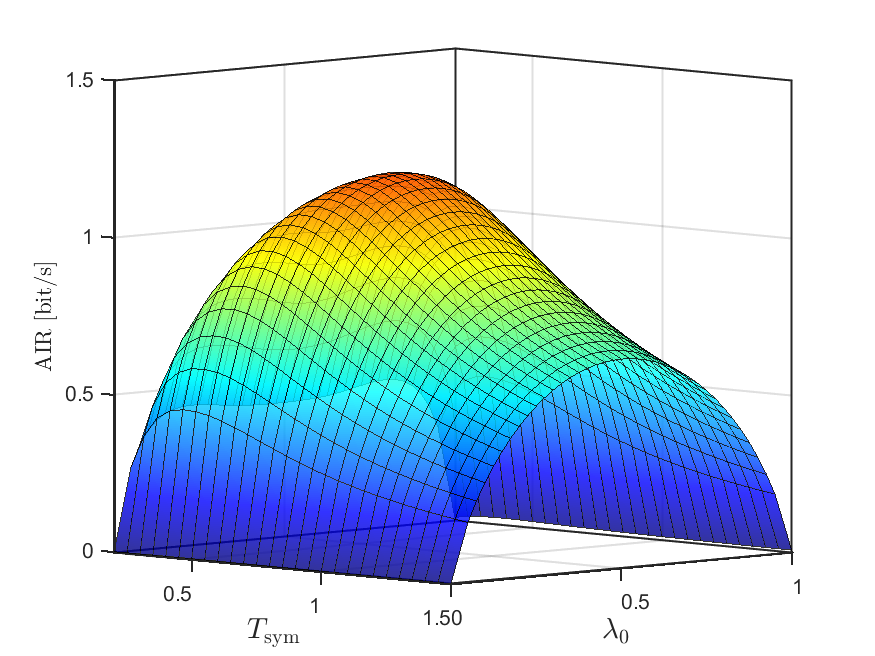}
    \caption{AIR as a function of the independent source probability of transmitting~$``0"$, $\lambda_0$, and of the STI $T_{\mathrm{sym}}$ in ISI-Unaware scenario.}
    \label{fig:IND_UNAWARE}
\end{figure}

 Figure~\ref{fig:C_ISIU_CRR_Tsym3e-1}\footnote{Please note that in Figs.~\ref{fig:C_ISIA_CRR_Tsym3e-1}, \ref{fig:C_ISIU_CRR_Tsym3e-1}, \ref{fig:C_ISIA_CRR_Tsym7e-1}, \ref{fig:C_ISIU_CRR_Tsym7e-1}, \ref{fig:IND_AWARE}, and \ref{fig:IND_UNAWARE} a deliberately consistent color spectrum indicating the range of AIR from $0$ to $1.5$ was employed. This choice was made to facilitate enhanced visual comparison between the figures, promoting reader comprehension.} displays the AIR in the context of ISI unawareness, where symbols are generated by a correlated source, with $T_{\mathrm{sym}}$$\,=\,$$0.3$\,s. The capacity achieved in this scenario is $C_{\text{ISIU}}^{\text{CRR}}$$\,=\,$$0.82$, with the correlated source input distribution of $p$$\,=\,$$0.35$ and $q$$\,=\,$$0.75$.

An intriguing observation is that in this case, the capacity is not attained with a stationary equiprobable distribution. Rather, it is achieved when the probability of having a $``0"$ after a $``1"$, and the likelihood of generating successive $``0"$ symbols, is higher compared to generating $``1"$. This optimal strategy can be readily understood since there is no knowledge of ISI, and with an STI of $T_{\mathrm{sym}}$$\,=\,$$0.3$\,s, where ISI is prominent, it is preferable to avoid generating $``1"$s.

Another noteworthy observation in the figure is that when $p$$\,\approx\,$$0.7$ and $q$$\,\approx\,$$0.3$, a similar AIR value to the capacity is observed. This region of the correlated source input space signifies the preference for generating more $``1"$s. In comparison to the ISI-Aware case shown in Fig.~\ref{fig:C_ISIA_CRR_Tsym3e-1}, we can observe that the AIR is lower in the ISI-Unaware scenario as expected but with the same color pattern.
\par Figure~\ref{fig:C_ISIA_CRR_Tsym7e-1} depicts the AIR for the same scenario as shown in Fig.~\ref{fig:C_ISIA_CRR_Tsym3e-1}, but here the STI is $T_{\mathrm{sym}}$$\,=\,$$0.7$\,s. In this case, the capacity is achieved with $p$$\,=\,$$q$$\,=\,$$0.55$, indicating a preference for almost equiprobable input distributions.

By comparing the capacity-achieving input distribution in Figs.~\ref{fig:C_ISIA_CRR_Tsym3e-1} and~\ref{fig:C_ISIA_CRR_Tsym7e-1} we can observe that the optimal values of $p$ and $q$ converge to $0.5$ as the STI increases. With $p$ and $q$ closer to $0.5$, the source behaves more similarly to a source that emits independent and uniformly distributed symbols.
\par Figure~\ref{fig:C_ISIU_CRR_Tsym7e-1} presents the AIR for a similar scenario as depicted in Fig.~\ref{fig:C_ISIU_CRR_Tsym3e-1}, with STI $T_{\mathrm{sym}}$$\,=\,$$0.7$\,s. The capacity is obtained when the input distribution is equiprobable, characterized by $p$$\,=\,$$q$$\,=\,$$0.57$, indicating a preference for equal probabilities from a stationarity perspective.

Comparing this capacity with the one shown in Fig.~\ref{fig:C_ISIU_CRR_Tsym3e-1}, we observe a shift in the optimal input distribution. In this case, the capacity is achieved with an equiprobable distribution that slightly avoids generating the same symbols successively, whereas in Fig.~\ref{fig:C_ISIU_CRR_Tsym3e-1}, the avoidance of generating $``1"$s was a little more preferred. This observation can be explained by the increased STI $T_{\mathrm{sym}}$, which leads to a relatively reduced impact of ISI.
\begin{figure}
    \centering
    \includegraphics[width=1\linewidth]{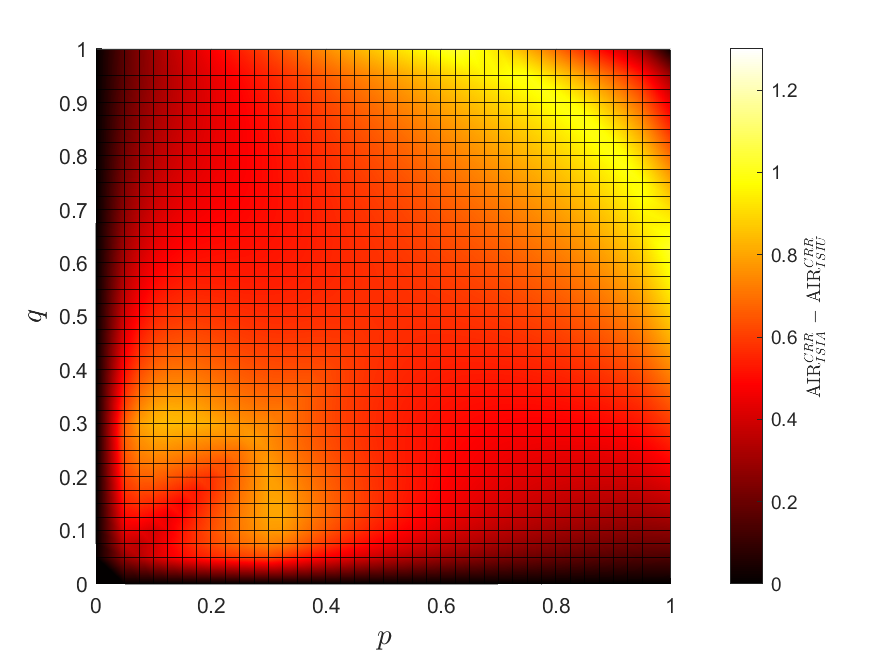}
    \caption{Difference between the AIR values corresponding to Fig.~\ref{fig:C_ISIA_CRR_Tsym3e-1} and Fig.~\ref{fig:C_ISIU_CRR_Tsym3e-1} over the input distribution space.}
    \label{fig:Diff_C_ISI_CRR_Tsym3e-1}
\end{figure}
\par In Fig.~\ref{fig:Diff_C_ISI_CRR_Tsym3e-1}, we present the difference between the AIRs illustrated in Fig.~\ref{fig:C_ISIA_CRR_Tsym3e-1} and Fig.~\ref{fig:C_ISIU_CRR_Tsym3e-1} to analyze the disparity between the ISI-Aware scenario and the ISI-Unaware case for an STI of $T_{\mathrm{sym}}$$\,=\,$$0.3$\,s. Consistent with the theoretical prediction stated in~\eqref{eq:ISIA<ISIU}, the difference between the two AIRs is non-negative.

There are two prominent regions where the difference between the AIRs is significant. The first region, located in the upper right side of the figure, demonstrates that the performance of the ISI-Unaware scenario tends to approach~$0$, whereas the ISI-Aware case exhibits a higher AIR in that region. A similar observation can be made for the bottom left region of the figure.
\par Since the independent source's input space can be spanned by a single variable, $\lambda_{0}$, we can examine the AIR values of the independent source as a function of the STI, $T_{\mathrm{sym}}$, and the probability of transmitting a $``0"$, $\lambda_{0}$. Figure~\ref{fig:IND_AWARE} illustrates the AIR values in the ISI-Aware scenario, where the source generates symbols independently, and the receiver is aware of the previously transmitted symbols. As we also observed in Fig.~\ref{fig:Capacity}, the highest capacity is achieved when $T_{\mathrm{sym}}=0.45$\,s.

\begin{figure}
    \centering
    \includegraphics[width=1\linewidth]{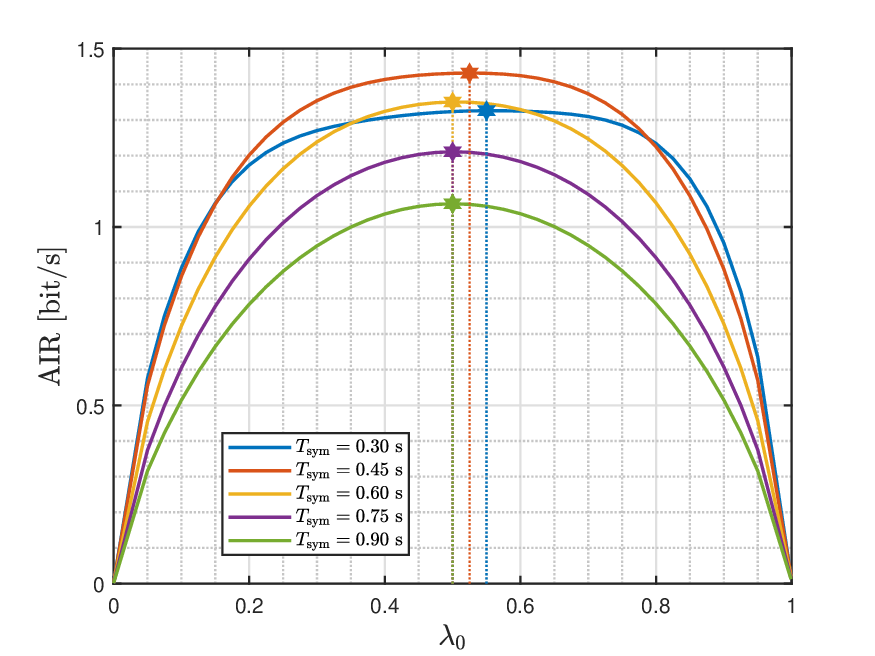}
    \caption{AIR values corresponding to $T_{\mathrm{sym}}$$\,\in\,$$\{0.30,0.45,0.60,0.75,0.90\}$\,s when the source is of the independent type, and ISI-Aware scenario holds. Hexagram markers indicate the capacity associated with each $T_{\mathrm{sym}}$.}
    \label{fig:Cross_AIR_ISIA_Tsym}
\end{figure}

To gain further insight, in Fig.~\ref{fig:Cross_AIR_ISIA_Tsym} we present a cross-sectional view of Fig.~\ref{fig:IND_AWARE}, focusing on specific STIs ($T_{\mathrm{sym}}$$\,\in\,$$\{0.30,0.45,0.60,0.75,0.90\}$\,s). The hexagram markers indicate the capacity corresponding to each STI. It is evident that as $T_{\mathrm{sym}}$ increases, the input distribution associated with the channel capacity gradually approaches an equiprobable input distribution (\textit{i.e., $\lambda_{0}$$\,=\,$$0.5$}). However, the maximum capacity does not occur when symbols are transmitted with equal probability. In fact, our analysis demonstrates a preference for transmitting a slightly higher number of $``0"$ symbols compared to $``1"$s. Interestingly, when $T_{\mathrm{sym}}$$\,=\,$$0.3$\,s, even for $\lambda_{0}$$\,<\,$$0.35$ and $\lambda_{0}$$\,>\,$$0.65$, we observe favorable AIR values compared to those associated with $T_{\mathrm{sym}}$$\,\geq\,$$ 0.6$\,s.
\par
\begin{figure}
    \centering
    \includegraphics[width=1\linewidth]{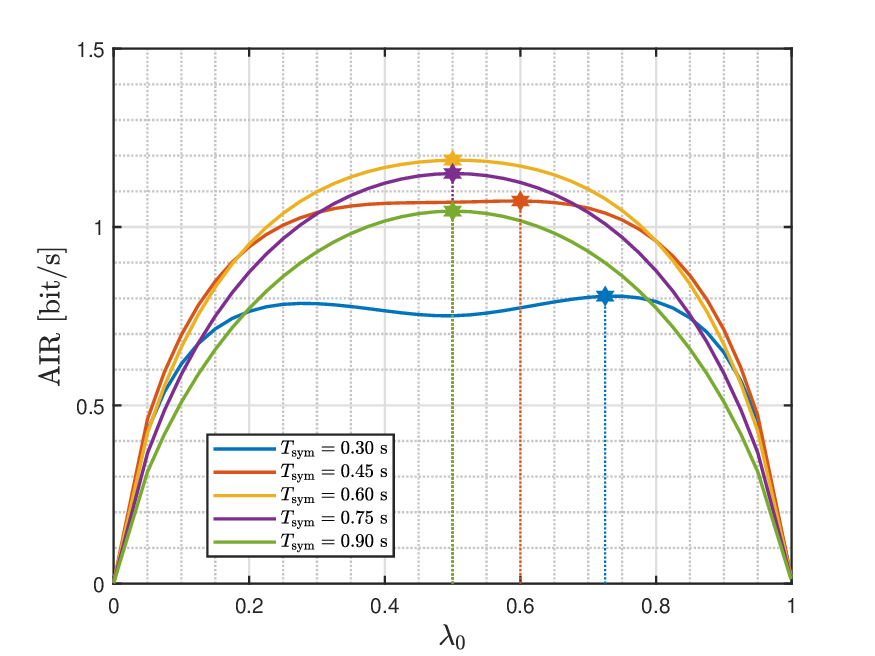}
    \caption{AIR values corresponding to $T_{\mathrm{sym}}$$\,\in\,$$\{0.30,0.45,0.60,0.75,0.90\}$\,s when the source is of the independent type, and ISI-Unaware scenario holds. Hexagram markers indicate the capacity associated with each $T_{\mathrm{sym}}$.}
    \label{fig:Cross_IND_UNAWARE}
\end{figure}

 Figure~\ref{fig:IND_UNAWARE} shows the AIR in a similar fashion as in Fig.~\ref{fig:IND_AWARE}, but assumes no knowledge on ISI. Compared to Fig.~\ref{fig:IND_AWARE}, we can observe that the AIR drops. However, the shape of the manifold remains similar. The cross-sectional view of Fig.~\ref{fig:IND_UNAWARE} is depicted in Fig.~\ref{fig:Cross_IND_UNAWARE} for the same set of $T_{\mathrm{sym}}$ as for Fig.~\ref{fig:Cross_AIR_ISIA_Tsym}. The maximum possible capacity achieved with equiprobability of input symbols is at $T_{\mathrm{sym}}$$\,=\,$$0.6$\,s. The reason for the complex shape of the AIR curves in Fig.~\ref{fig:Cross_AIR_ISIA_Tsym} is not trivial: each $\lambda_0$ is associated with a different channel,  which depends on the specific optimum detector threshold,~$\tau$. The curve for $T_{\mathrm{sym}}$$\,=\,$$0.3$\,s exhibits two local maxima. The maximum at $\lambda_{0}$$\,\approx\,$$0.28$ suggests transmitting fewer $``0"$s is beneficial, which may seem counter-intuitive given the higher ISI associated with faster transmission rates. However, the other maximum at $\lambda_{0}$$\,\approx\,$$0.75$ suggests transmitting more $``0"$s is optimal. This observation is sensible because the ISI increases with the transmission rate. By transmitting $``1"$ less frequently, the ISI is reduced, yielding an improvement in the AIR. As expected, the maximum associated with $\lambda_{0}$$\,\approx\,$$0.75$ is higher than the one associated with~$\lambda_{0}$$\,\approx\,$$0.28$.
\section{Conclusions}\label{sec:conclusion}

We have investigated the Achievable Information Rate (AIR) of a diffusive molecular communication (MC) channel with a fully absorbing receiver, which counts particles absorbed along each symbol time interval (STI) and resets the counter at every interval. The MC channel is affected by memory and thus inter-symbol interference (ISI), due to the delayed arrival of molecules. 
To reduce the complexity in calculating the mutual information (MI), we have measured the effective memory length as an integer number of STIs and considered a single-symbol memoryless detector. Unlike previous works, we have also optimized the detector threshold to MI. We have approximated as Gaussian the received signal distribution and calculated the channel MI affected by ISI. Our investigation on AIR covers four distinct scenarios as the independent source and correlated source with and without knowledge about the previously transmitted symbols at the receiver side.

Our selection of numerical results demonstrates that, in general, with correlated source, we can achieve higher capacity. The optimal input probability distribution achieving the capacity may not be uniform. In particular, when the STI $T_{\mathrm{sym}}$ is small, thus implying strong ISI, the maximum AIR does not occur with the equiprobable transmission of symbols.
\appendices
  \section{}
  In this section, we obtain an equivalent expression to describe the probability of an event conditioned on the presence of two other joint events that was extensively used in the manuscript.
  \label{FirstAppendix}
  \begin{align}
    \begin{split}
        P_{A|B,C}(a|b,c) &= \frac{P_{C|A,B}(c|a,b)P_{A,B}(a,b)}{P_{C,B}(c,b)}\\& =\frac{P_{C|A,B}(c|a,b)P_{A|B}(a|b)P_{B}(b)}{P_{C|B}(c|b)P_{B}(b)}\\&
        =\frac{P_{C|A,B}(c|a,b)P_{A|B}(a|b)}{P_{C|B}(c|b)}\\&
        = \frac{P_{C|A,B}(c|a,b)P_{A|B}(a|b)}{\sum\limits_{\substack{\forall y} } P_{C,A|B}(c,y|b)}\\&
        =\frac{P_{C|A,B}(c|a,b)P_{A|B}(a|b)}{\sum\limits_{\substack{\forall y} }P_{C|A,B}(c|y,b)P_{A|B}(y|b)}
    \end{split}
\end{align}
\ifCLASSOPTIONcaptionsoff
  \newpage
\fi
\bibliographystyle{IEEEtran}
\bibliography{Bibliography}

\end{document}